\documentclass[a4paper,11pt]{article}
\pdfoutput=1 % if your are submitting a pdflatex (i.e. if you have
\usepackage{jcappub} % for details on the use of the package, please
\usepackage[T1]{fontenc} % if needed
\usepackage{amsmath}
\usepackage{amsfonts,txfonts,pxfonts}
\usepackage{amssymb}
\usepackage{orcidlink}
\usepackage[export]{adjustbox}
\usepackage{fix-cm}
\usepackage{adjustbox}
\usepackage{latexsym,amsmath,amssymb}
\RequirePackage{graphicx}
\RequirePackage{mathptmx}
\usepackage[arrowdel]{physics}
\usepackage{graphicx,epstopdf}
\usepackage{slashed}
\usepackage{bm}
\usepackage{cases}
\usepackage{color}
\usepackage{bm}
\usepackage{dcolumn}
\usepackage{lipsum}
\usepackage{hyperref}
\usepackage{lineno}
\arxivnumber{} % Only if you have one
\title{Cosmological dynamics and structure formation in a generalized mass-to-horizon entropy-inspired modified gravity}
\author{Subhra Mondal\orcidlink{0009-0003-6469-6238}}
\author{and Amitava Choudhuri\orcidlink{0000-0001-9499-8585}}
\affiliation{Department of Physics,
 \href{https://ror.org/05cyd8v32}{The University of Burdwan}, Golapbag, Purba Bardhaman - 713104, West Bengal, India.}
\emailAdd{cosmology313@gmail.com}
\emailAdd{amitava\_ch26@yahoo.com}
\abstract{In this article, our goal is to investigate the cosmological dynamics and structure formation in a modified cosmological framework inspired by a generalized mass-to-horizon entropy relation and consistent with the Clausius relation. Invoking the gravity-thermodynamics conjecture leads to alterations in the Friedmann equations as well as Hubble parameter evolution. The effects of the generalized entropy on various cosmographic parameters and on the growth of the linear matter perturbations by constructing perturbed field equations via employing spherical collapse formalism in a flat Friedmann-Lema\^{i}tre-Robertson-Walker background have been explored. We discuss a novel and well-known diagnostic approach to differentiate various cosmological models vis-à-vis flat and non-flat $\Lambda$CDM frameworks, and find that the generalized mass-to-horizon entropy-inspired modified cosmology ($n\ne 1$) successfully passes all the litmus tests by falsifying both the flat and non-flat $\Lambda$CDM paradigms. It is shown that this model also satisfies the requirements for the Universe to achieve thermodynamic equilibrium in the distant future. We also study the halo mass function and cluster number counts in this modified gravity scenario. All the results are compared with the fiducial $\Lambda$CDM profile, showing that the additional entropic correction influences the expansion history, the growth rate of structures, and the abundance of collapsed halos. We observe that the more massive collapsed structures are less abundant and form at later epochs, which is expected from the hierarchical model of large-scale structure formation.}
\keywords{generalized mass-to-horizon entropy, modified gravity, cosmography, top-hat spherical collapse, halo mass function, structure formation} % Add your keywords here
\begin{document}
\maketitle
\flushbottom
%%%%%%%%%%%%%%%%%%%%%%%%%%%%%%%%%%%%%%%%%%%%%%%%%%%%%%%%%%%%%%%%%%%%%%%%%%%%%%%%%%%%%%%%%%%%%%%%%%%%%%%%%%%%%%%%%%%%%%%%%%%%%%%%%%%%%%%%%%%%%%%
\section{Introduction}
\par General Relativity (GR) alongside Quantum Field Theory (QFT) is a cornerstone of contemporary physics and has proven extremely successful \cite{carroll2019spacetime}. The cosmology associated with GR, namely the $\Lambda$CDM standard model, has rendered exceptional achievement in explaining several facets of cosmology, from the power spectrum and statistical properties of Cosmic Microwave Background Radiation anisotropies \cite{page2003first} to the late-time accelerated Universe \cite{perlmutter1999measurements,riess1998observational}. This model also efficiently captures the features of large-scale cosmological structures in the Universe \cite{bernardeau2002large,bull2016beyond} while remaining consistent with the abundance of light elements such as helium, hydrogen, and others \cite{schramm1998big,steigman2007primordial,cyburt2016big,iocco2009primordial}. Despite its straightforwardness and the impressive success of the $\Lambda$CDM model, its validity is currently under intense observation due to several substantial theoretical and observational challenges faced by this model \cite{abdullah2022coherent,anchordoqui2021dissecting,buchert2016observational,schmitz2022modern,di2021realm}. Consequently, alternative theories of gravity/cosmology have been proposed to account for the observed phenomena.
\par Primarily, there are a few theoretical strategies to develop a theory beyond GR or the $\Lambda$CDM paradigm. The first strategy focuses on altering the geometry of gravity. This is explored through extensions of the Einstein-Hilbert action, rather than strictly following Einstein’s original GR formulation, which results in a diverse array of models grouped under the term modified gravity theories \cite{nojiri2017modified, capozziello2011extended}. A different strategic approach that maintains GR as the foundational theory of gravity while modifying the matter sector. In this method, the inclusion of unprecedented dynamical quantities, such as inflaton fields or DE-like fluids, plays a pivotal role in facilitating cosmic acceleration \cite{Olive1989inflation, guth1981inflationary,copeland2006dynamics,sahni2006reconstructing,li2011dark}.  
\par An alternative and equally important viewpoint suggests a deep inner-connection between thermodynamics and gravity \cite{padmanabhan2005gravity,jacobson1995thermodynamics,eling2006nonequilibrium,padmanabhan2006gravity,kothawala2007einstein,padmanabhan2012dialogue,padmanabhan2002classical,nojiri2025modified}. The first law of thermodynamics can be applied to this boundary to derive the governing field equations in this formulation, which views the Universe as a thermodynamic system confined by the apparent horizon \cite{akbar2006friedmann,akbar2007thermodynamic,cai2007unified,cai2005first}. As long as the corresponding area laws are suitably generalized, this method can be applied to a variety of modified gravity scenarios in addition to conventional GR \cite{paranjape2006thermodynamic,akbar2006friedmann,cai2010horizon,jamil2010thermodynamics}. It is nothing new to say that horizon entropy gets special attention in the context of entropic cosmology, which arises from generalizing black hole (BH) thermodynamics on cosmological scales, the holographic principle, and the entropic force. The cosmic acceleration is the natural consequence of the entropic force, which arises from incorporating the Gibbon-Hawking-York (GHY) surface term to the usual Einstein-Hilbert action \cite{easson2011entropic,easson2012entropic}. The new action for the gravitational field with matter fields over a region
$\mathcal{V}$ with boundary $\partial\mathcal{V}$ has the form\footnote{Throughout this article, the lowercase Latin indices \(a, b, \ldots, i, j, \ldots\) represent the values \(0, 1, 2, 3\) unless stated otherwise. The index \(0\) signifies the time dimension, while the indices \(1, 2, \) and \(3\) relate to the typical spatial dimensions. We use the $(-,+,+,+)$ format of spacetime signature. Standard notations are used for the universal constants, including the speed of light \(c\), Planck constant \(\hbar\), Boltzmann constant \(k_B\), and gravitational constant \(G\).}
\begin{equation}
I = \underbrace{\int_{\mathcal{V}} \left( \frac{c^4 \mathcal{R}}{16\pi G} + \mathcal{L}_{matter} \right) \sqrt{\mathrm{-g}} \, d^4x}_{\text{Einstein-Hilbert + Matter terms }} + \underbrace{\frac{c^4}{8\pi G} \oint_{\partial\mathcal{V}} \mathcal{K} \sqrt{{\mathfrak{h}}} \, d^3y}_{\text{Gibbons-Hawking-York term}}
\end{equation}
where $\mathcal{R}$ is denoted as the Ricci scalar, $\mathcal{L}_{matter}$ is the matter Lagrangian, $\mathrm{g}$ is the determinant of the covariant components of bulk metric tensor $g_{ab}$, $\mathcal{K}$ is the trace of the extrinsic curvature of the boundary, and $\mathfrak{h}$ is the determinant of the induced metric $\mathfrak{h}_{ab}$ lives on the boundary \cite{hawking1996gravitational,gibbons1977action,basilakos2012generalizing}. The GHY surface term in the new action plays an important role in describing the thermodynamic properties of the cosmological horizon and leads to modified Einstein’s standard field equations \cite{easson2011entropic,easson2012entropic,mondal2024dynamics,mondal2025dynamics}. In this context, it is important to highlight the fact that entropic cosmology is developed within the extended GR formalism in the FLRW background. This differs from  Padmanabhan's emergent gravity \cite{padmanabhan2012emergence} and Verlinde’s entropic gravity \cite{verlinde2011origin} frameworks. Although they originate from different contexts, both approaches interpret gravity as an emergent phenomenon rather than a fundamental interaction, offering a radical departure from geometric descriptions.
\par Throughout the years, numerous generalized entropy measures have been developed as extensions of the semiclassical Bekenstein-Hawking entropy and have become a prime focus of researchers. These measures emerge from nonstandard statistical mechanics or from considerations related to quantum physics and gravity on the holographic horizon. Some popular examples include Tsallis \cite{tsallis2009introduction,tsallis1988possible}, R\'{e}nyi \cite{renyi1961measures}, and Sharma-Mittal \cite{sharma1975new} entropies, which possess non-extensivity; Kaniadakis \cite{kaniadakis2002statistical,kaniadakis2001non} entropy, which is based on relativistic statistical mechanics; and Barrow \cite{barrow2020area} entropy, which considers quantum-gravity corrections to horizon geometry. These formulations retrieve the standard Bekenstein-Hawking entropy under certain conditions on the model parameters. As a consequence, encompassing them in the thermodynamic framework of gravity has initiated considerable interest nowadays \cite{tsallis2013black,tsallis2019black,d2024lagrangian,barrow2020area,saridakis2020modified,mondal2024dynamics,mondal2025dynamics,basilakos2025modified,gohar2024generalized,nojiri2025modified,luciano2026new,ghaffari2019black,lymperis2021modified,sheykhi2010thermodynamics,sheykhi2011power,sheykhi2010entropic,cai2008corrected}.
\par A criticism centered on cosmology in Refs. \cite{gohar2024foundations,gohar2024generalized,sheykhi2025emergence} emphasizes an important limitation in establishing models based on entropic gravity and holographic principle scenarios. They provide an argument that demands that when a couple of conditions: (a) the temperature of the cosmic horizon maintains thermodynamic consistency as defined by the Clausius relation, and (b) the mass-to-horizon relation follows a linear pattern, are fulfilled, the resultant cosmological model becomes indistinguishable from the conventional one derived from Bekenstein-Hawking entropy. This compels all such cosmological models to exhibit the same flaws, including their failure to accurately align with the observed cosmological dynamics at both the background and perturbative levels \cite{basilakos2012generalizing,basilakos2014entropic}. In order to overcome these inherent flaws, a generalized mass-to-horizon relation has been put forward in Ref. \cite{gohar2024generalized}, which naturally leads to a revised form of entropy that includes recently proposed forms such as Tsallis-Cirto \cite{tsallis2013black,tsallis2019blackholeentropy}, Tsallis-Zamora \cite{zamora2022thermodynamically1,zamora2022thermodynamically2} and Barrow \cite{barrow2020area} entropies as specific examples.
\par Recent studies on cosmological implications of the \textit{generalized mass-to-horizon entropy} (GMHE) framework have been explored in different articles \cite{gohar2024generalized,luciano2025modified,luciano2025modified_Cos,basilakos2025modified,luciano2026baryogenesis,sheykhi2025emergence, Mondal2026Gravitational}. An investigation in Ref. \cite{gohar2024generalized} shows that this framework can produce a cosmological model that fits observational data similarly to the standard $\Lambda$CDM model under certain conditions on model parameters. The gravitational origin of the GMHE through the Iyer-Wald Lagrangian approach has been examined in Ref. \cite{Mondal2026Gravitational} by suggesting a form of $f(\mathcal{R})$ gravity that successfully replicates this generalized horizon entropy. The article in Ref. \cite{sheykhi2025emergence} derives modified Friedmann equations by employing mass-to-horizon entropy and examines the origin of cosmic space by merging the equilibrium approach, motivated by the first law of thermodynamics applying to the apparent horizon, and Padmanabhan's idea of the emergence of cosmic space with time. Utilizing the gravity-thermodynamics conjecture, authors in Ref. \cite{basilakos2025modified} extract the modified Friedmann equations from this entropy and incorporate an effective DE component from additional terms in the entropy formula. Their analysis with different datasets shows that this modified model is in agreement with observations. The generation of the baryon asymmetry within this cosmological framework has been studied in Ref. \cite{luciano2026baryogenesis} and predicts bounds on the entropic parameters. In a study in Ref. \cite{luciano2025modified}, observational constraints were placed on this model, and observe this model shows no significant deviation from the $\Lambda$CDM profile. Additionally, the impact of this modified cosmology on matter perturbation growth and primordial gravitational waves in the early Universe has been investigated in Ref. \cite{luciano2025modified_Cos}.
\par Current cosmological observation of distant Supernovae Type Ia provides data that confirms that the Universe is made of roughly 5\% visible baryonic matter, 25\% dark matter (DM), and 70\% dark energy (DE) \cite{perlmutter1999measurements,de2000flat,yang2020evidence}. There is a substantial amount of proof reinforcing the existence of invisible DM and DE in our Universe \cite{seljak2006cosmological}. The DE has earned attention recently and is supposed to be a mysterious component that exerts negative pressure, causing the Universe's expansion to accelerate by showing its anti-gravity nature. Another mysterious component, DM, plays a couple of crucial roles in the evolution of the Universe. Firstly, it provides the necessary gravitational attraction for the rotation of spiral galaxies and galaxy clusters. Secondly, the perturbations need to grow in the early Universe that form the currently observed cosmic structures. From the epoch of matter-domination until decoupling, perturbations in DM can grow because DM does not interact through electromagnetic forces like baryonic matter does, allowing it to collapse under its own gravitational instability; however, perturbations in baryonic matter cannot, since they are tightly coupled to photons through Thomson scattering. After decoupling, the baryons become free from the pressure support supplied by the photon radiation, and they fall into the DM potential wells in a short duration. Within a few expansion times after decoupling, the baryonic perturbations catch up with the DM perturbations. The baryonic matter follows the distribution of DM due to gravitational attraction. Consequently, galaxy clusters are embedded within the halos of DM. Hence, the observed arrangement of galaxy clusters offers insights into the distribution of DM halos throughout the Universe. If DM were absent, the formation of galaxies would have occurred much later than we currently observe \cite{kolb2018early,farsi2022structure}.
\par The formation of large-scale structures in the Universe is a fascinating as well as complex problem in cosmology. It tells how our Universe evolved from a uniform and smooth state to the highly clustered form we see today. It all initiated during the inflationary phase when small quantum disturbances in the scalar curvature served as the building blocks for large-scale cosmic structures. As the Universe expanded rapidly during this inflation, these small fluctuations were amplified, leading to gravitational instabilities that eventually shaped the galaxies and clusters we observe today. In essence, the collapsed regions that formed in the early Universe served as the initial cosmic seeds for density perturbations from which the vast structures developed in the late-time Universe \cite{kolb2018early,peebles2020principles,white1978core}.
\par The most straightforward and suitable (semi-)analytical procedure to study the evolution of perturbations of in-falling masses into a bound system and structure formation is the top-hat spherical collapse (SC) model \cite{gunn1972infall,abramo2007structure}. It has been confirmed that the SC equations can actually be derived from GR, provided shear is not a significant factor \cite{gaztanaga2001nonlinear}. In this classical top-hat model, one analyzes a uniform and symmetric spherical perturbation throughout the perturbed region within a homogeneous background Universe \cite{fernandes2012spherical}. The symmetry inherent in this model encourages us to address a spherical perturbation in a Friedmann-Lema\^{i}tre-Robertson-Walker (FLRW) Universe. Effectively, we can describe the growth of perturbations in a spherical region using the same Friedmann equations, but a different scale factor that governs the underlying gravitational theory \cite{planelles2015large}. The SC model indicates that during the early stages, primordial spherical overdense regions expand in accordance with the Hubble flow. The mechanism of collapse resulting from gravitational instability is highly dependent on the dynamics of the background Hubble flow at the early epoch \cite{naderi2015evolution}. As the relative overdensity of spherical regions compared to the background remains small, linear theory is sufficient for studying their evolution dynamics. At a specific moment, gravity begins to dominate and exceed the rate of expansion. Ultimately, the overdense sphere reaches its maximum size and completely detaches from the background expansion. The following phase is characterized by the collapse of the spherical region due to its own self-gravity.
\par In the study of the evolution of matter density perturbations and structure formation, numerous works have been published across various cosmological frameworks. These include Rastall gravity \cite{ziaie2020structure}, Dvali, Gabadadze, and Porrati (DGP) braneworld cosmology \cite{mukherjee2020spherical}, reconstructed dark energy models \cite{mukherjee2025spherical}, $\Lambda$ viscous cold dark matter (CDM) Universe \cite{velten2014structure}, mimetic gravity \cite{farsi2022structure}, energy-momentum squared gravity \cite{farsi2023evolution}, and others \cite{brax2012structure,koyama2006structure}. In Ref. \cite{abramo2007structure}, non-linear structure formation accompanied by dark energy perturbations has been studied. In one of our previous works in Ref. \cite{mondal2024temporal}, the cosmological evolution of density perturbations of the Bose–Einstein condensate DM modeled by Gross–Pitaevskii–Poisson (GPP) system via Jeans' instability has been investigated.
\par Here in this article, we aim to investigate the cosmological dynamics and structure formation in a modified cosmological framework inspired by a nonlinear GMHE relation and consistent with the Clausius relation, intending to identify a potential candidate that could distinguish this model from the traditional $\Lambda$CDM profile and other alternative gravity scenarios as well. Specifically, we investigate the effects of entropic modifications on the non-perturbative as well as perturbative regimes during the late times of the Universe, focusing on the linear growth of structures and the abundance of collapsed halos. These attributes show a distinction from the earlier work presented in Ref. \cite{luciano2025modified}, where perturbations are studied only up to the matter-dominated epoch.  
\par The present article is organized as follows. In Section \ref{GMHE-inspired modified cosmology}, we present a brief review of GMHE-inspired modified cosmology and discuss its dynamics. In Sec. \ref{Growth of matter spherical overdensities in GMHE-inspired modified cosmology}, we explore linear growth of matter inhomogeneities in the spherical top-hat collapse approach of spatially flat GMHE-inspired modified cosmology. For the same cosmological setup, we study the halo mass function and number counts of the collapsed objects in Sec. \ref{Halo mass function and cluster number counts in GMHE-inspired modified cosmology} using the Sheth-Mo-Tormen method. Section \ref{Discussion and Conclusions} is devoted to discussion and conclusions.
%%%%%%%%%%%%%%%%%%%%%%%%%%%%%%%%%%%%%%%%%%%%%%%%%%%%%%%%%%%%%%%%%%%%%%%%%%%%%%%%%%%%%%%%%%%%%%%%%%%%%%%%%%%%%%%%%%%%%%%%%%%%%%%%%
\section{GMHE-inspired modified cosmology}\label{GMHE-inspired modified cosmology}
\par Let us begin with a $(1+3)$-dimensional homogeneous and isotropic FLRW Universe following metric \cite{bak2000cosmic,cai2005first}
\begin{equation}\label{metric}
ds^{2} = h_{ab}\, dX^{a} dX^{b} + \tilde{r}^{\,2}\, d\Omega_{2}^{2} ,
\end{equation}
where \( \tilde{r} = a(t)\, r \) and \( X^{0} = ct,\; X^{1} = r \). The $2$-dimensional metric is delineated as \( h_{ab} = \mathrm{diag}(-c^2,\; a^{2}(t)/(1 - \kappa r^{2})) \)
where \( \kappa \, = +1,\, 0,\, \text{and}\, -1 \)
is the spatial curvature constant for spherical, Euclidean, and hyperbolic Universes, respectively, and \(d\Omega_{2}^{\,2} = d\theta^2 + \sin^2{\theta}\,d\phi^2\) denotes the line element of the $2$-dimentional unit sphere. Here, the scale factor \(a \equiv a(t)\) characterizes the cosmological expansion of the Universe. For simplicity, we assume the components of the Universe, such as non-relativistic pressureless dust matter (visible matter + CDM), DE (neglecting the radiation at lower redshifts), are modeled as a perfect fluid, with the covariant energy-momentum tensor components \cite{carroll2019spacetime,mondal2024dynamics,mondal2025dynamics}
\begin{equation}\label{EMT}
    \mathcal{T}_{ab}= \left(\rho + \frac{P}{c^2}\right)u_a u_b + P g_{ab}\,,
\end{equation}
where $\rho$ and $P$ indicate the density and pressure of the fluid, respectively. The four-velocity $1$-form components of the fluid $u_a$ follow the normalization condition $u_a u^b=-1$. In a curved space-time manifold, $\nabla_a\mathcal{T}^{ab}=0$ characterizes the influence of external gravitational forces on the fluid, leading to the continuity equation \cite{carroll2019spacetime,mondal2024dynamics,mondal2025dynamics}
\begin{equation}\label{continuity_equation}
    \dot{\rho} + 3H\left(\rho + \frac{P}{c^2}\right)=0\,\,.
\end{equation} 
Further, we consider our Universe is bounded by a physical boundary called the dynamical apparent horizon, a marginally entrapped surface with vanishing expansion, determined by the relation \( h_{ab}\, \partial^{a}\tilde{r}\, \partial^{b}\tilde{r} = 0 \), having a radius \cite{bak2000cosmic,cai2005first,akbar2007thermodynamic,sheykhi2007thermodynamical,mondal2024dynamics,mondal2025dynamics}
\begin{equation}\label{apparent_horizon}
    \tilde{r}_{hor} = \frac{c}{\sqrt{H^2 + \kappa c^2/a^2}}\,.
\end{equation}
Here, \(H \overset{\text{def}}{=} \dot{a}(t)/a(t)\) represents the Hubble parameter. The over-dots represent the order of derivatives with respect to the cosmic time $t$ throughout this article. For a dynamical FLRW horizon, the surface gravity is \cite{akbar2007thermodynamic}
\begin{equation}
\kappa_{SG} = \frac{1}{2\sqrt{-h}}\, \partial_a \left( \sqrt{-h}\, h^{ab}\, \partial_b \tilde r \right) = -\frac{c^{2}}{\tilde{r}_{hor}}
\left(1 - \frac{\dot{\tilde r}_{hor}}{2 H \tilde r_{hor}}\right)\, .
\end{equation}
The temperature associated with the apparent horizon can be described as \cite{akbar2007thermodynamic}
\begin{equation}\label{temp}
T_{hor} = \frac{\hbar\, |\kappa_{SG}|}{2\pi\, k_{B}\, c} = \frac{\hbar c}{2\pi k_{B}\tilde r_{hor}}\,\left|
1 - \frac{\dot{\tilde r}_{hor}}{2H\tilde r_{hor}}\right|\,.
\end{equation}
Interpreting the dynamic apparent horizon requires understanding the surface gravity, which not only relies on the Hubble parameter and the radius of the apparent horizon, but also on the time variation of the horizon radius. For a static BH, the surface gravity remains constant according to the zeroth law of BH thermodynamics \cite{bardeen1973four}. Even a little change in the mass of a BH will induce changes in the horizon radius, horizon entropy, and  Hawking temperature. However, the first law of thermodynamics, or the Clausius relation
\begin{equation}\label{Clausius_rel}
    -dE = c^2 dM = T_{hor}\,dS_{hor}\,,
\end{equation}
where \(E\), \(M\), and \(S\) represent energy, mass, and entropy associated with the cosmic horizon in connection with the holographic principle, does not compel us to consider this temperature change during its application. Thus, when applying the first law of thermodynamics to the apparent horizon to compute temperature through surface gravity, or considering an infinitesimal amount of energy crossing the apparent horizon, the radius \( \tilde{r}_{hor} \) can be assumed constant \cite{cai2005first}. In this sense, Eq. (\ref{temp}) is approximated to
\begin{equation}\label{temp_approx}
    T_{hor}= \frac{\hbar c}{2\pi k_{B}\tilde r_{hor}}\Longrightarrow  \text{Hawking temperature}\,.
\end{equation}
In particular, it is well-known that the Bekenstein-Hawking entropy \cite{bekenstein1973black,bekenstein2020black,bekenstein1974generalized,hawking1974black}
\begin{equation}
    S_{BH} = \frac{k_B\pi \tilde r\,^2_{hor}}{L^2_{Pl}}
\end{equation}
can be derived using the Hawking temperature (\ref{temp_approx}) and linear standard mass-to-horizon relation (SMHR)
\begin{equation}
    M = \frac{c^2}{G}\tilde r_{hor}
\end{equation}
in Clausius relation (\ref{Clausius_rel}).
Following the same approach, Gohar and Salzano \cite{gohar2024generalized} formulated a novel entropy for the cosmological horizon \cite{gohar2024generalized,luciano2025modified,luciano2025modified_Cos,basilakos2025modified,luciano2026baryogenesis,sheykhi2025emergence}
\begin{equation}\label{mod_entropy}
    S_{GS} =  \gamma\,\frac{2 \pi k_B}{L^2_{Pl}}\frac{ n}{n+1}\tilde r\,^{n+1}_{hor} = \gamma\,\frac{2n}{n+1}\tilde r\,^{n-1}_{hor}S_{HB}
\end{equation}
but by employing a generalized mass-to-horizon relation (GMHR) \cite{gohar2024generalized,luciano2025modified,basilakos2025modified,luciano2025modified_Cos,sheykhi2025emergence}
\begin{equation}\label{GMHR}
    M = \gamma\frac{c^2}{G}\tilde r^n_{hor}\, ,
\end{equation}
where mass may not scale linearly with horizon radius. Here, the entropic exponent parameter $n$ represents a non-negative real number, while the positive multiplicative parameter $\gamma$ possesses dimensions of $[L]^{1-n}$. Specifically for $\gamma =n =1$, the SMHR, as well as the standard Bekenstein-Hawking entropy $S_{BH}$, can be recovered. In this context, we mention that we now have enough flexibility to retrieve some other kinds of entropy apart from the standard case. For instance, if we consider $n = 2\delta - 1$, we obtain the nonextensive Tsallis-Cirto entropy \cite{tsallis2013black,tsallis2019blackholeentropy,mondal2024dynamics} and Tsallis-Zamora entropy \cite{zamora2022thermodynamically1,zamora2022thermodynamically2} when $n = d - 1$. Likewise, we recover quantum-corrected Barrow entropy \cite{barrow2020area,mondal2025dynamics} for $n = 1 + \Delta $\,.
\par According to Refs. \cite{cai2005first,gong2007friedmann}, the heat flux $\delta Q$ crossing through the apparent horizon, i.e., the change of the energy inside the apparent horizon $-dE$, can be expressed as
\begin{equation}\label{heat_flux}
   \delta Q = -dE = 4 \pi H (c^2\rho+P)\tilde r\,^3_{hor}\,dt\,.
\end{equation}
For a detailed discussion on the expressions of temperature and energy flux through the apparent horizon, one may see Ref. \cite{akbar2007thermodynamic}. Substituting Eqs. (\ref{temp_approx}), (\ref{mod_entropy}), and (\ref{heat_flux}) in Eq. (\ref{Clausius_rel}), one obtain the modified second Friedmann equation \cite{basilakos2025modified}
\begin{align} \label{2nd_Friedmann_eqn}
\left( H^{2} + \frac{\kappa c^2}{a^{2}} \right)^{\frac{1-n}{2}} \left( \dot{H} - \frac{\kappa c^2}{a^{2}} \right) &= -\frac{4 \pi G}{\gamma\, n} \left( \rho + \frac{P}{c^2} \right) = -\frac{4 \pi G}{\gamma\, n} \left( \rho_m + \rho_{\Lambda} + \frac{P_m +P_{\Lambda}}{c^2} \right) \,.
\end{align}
In evaluating this, we also exploit the relation 
$\dot{\tilde r}_{hor} = - H \tilde r\,^3_{hor}\left(\dot{H} - \frac{\kappa c^2}{a^2}\right)$. 
Here, $\rho_m$, $\rho_{\Lambda}= \frac{\Lambda}{8\pi G}$, $P_m = 0$ and $P_{\Lambda}= -c^2 \rho_{\Lambda} = -\frac{\Lambda c^2}{8\pi G}$ are the densities of matter and DE that comprise the Universe. By using Eq. (\ref{continuity_equation}) and integrating Eq. (\ref{2nd_Friedmann_eqn}), we eventually get the first Friedmann equation in GMHE-inspired modified cosmology \cite{basilakos2025modified}
\begin{equation}\label{mod_Friedmann_eqn}
\left( H^{2}
+ \frac{\kappa c^2}{a^{2}}\right)^ \frac{3-n}{2}= \Gamma_{\gamma,n}\, \rho = \Gamma_{\gamma,n}\, (\rho_m + \rho_{\Lambda}) 
\end{equation}
where $\Gamma_{\gamma,n} = \frac{4\pi G (3-n)}{3\gamma\, n}$ is a constant depending on parameters $\gamma$ and $n$.
%%%%%%%%%%%%%%%%%%%%%%%%%%%%%%%%%%%%%%%%%%%%%%%%%%%%%%%%%%%%%%%%%%%%%%%%%%%%%%%%%%%%%%%%%%%%%%%%%%%%%%%%%%%%%%%%%
\par As usual, we define the cosmological density parameters for matter, DE, and spatial curvature as
\begin{equation}\label{density_param}
    \Omega^{mod}_m \overset{\text{def}}{=} \frac{\rho_m}{\rho^{mod}_{cr}}\,,\quad  \Omega^{mod}_\Lambda \overset{\text{def}}{=} \frac{\rho_\Lambda}{\rho^{mod}_{cr}}\,,\quad \Omega^{mod}_\kappa \overset{\text{def}}{=} \frac{\kappa c^2}{a^2 H^2}\,
\end{equation}
respectively, where $\rho^{mod}_{cr} = H^{3-n}/\Gamma_{\gamma,n}$ is the critical density in the GMHE-inspired modified Universe. Therefore, the first Friedmann equation (\ref{mod_Friedmann_eqn}) can be expressed in terms of cosmological density parameters as
\begin{equation}
    \Omega^{mod}_m + \Omega^{mod}_{\Lambda} = (1+\Omega^{mod}_\kappa)^\frac{3-n}{2}\, .
\end{equation}
For a flat Universe with $\kappa = 0$, the above relation takes the form
\begin{equation}
    \Omega^{mod}_m + \Omega^{mod}_{\Lambda} = 1\, .
\end{equation}
\begin{figure}[htbp!]
    \centering
    \includegraphics[width=10cm,height=8cm]{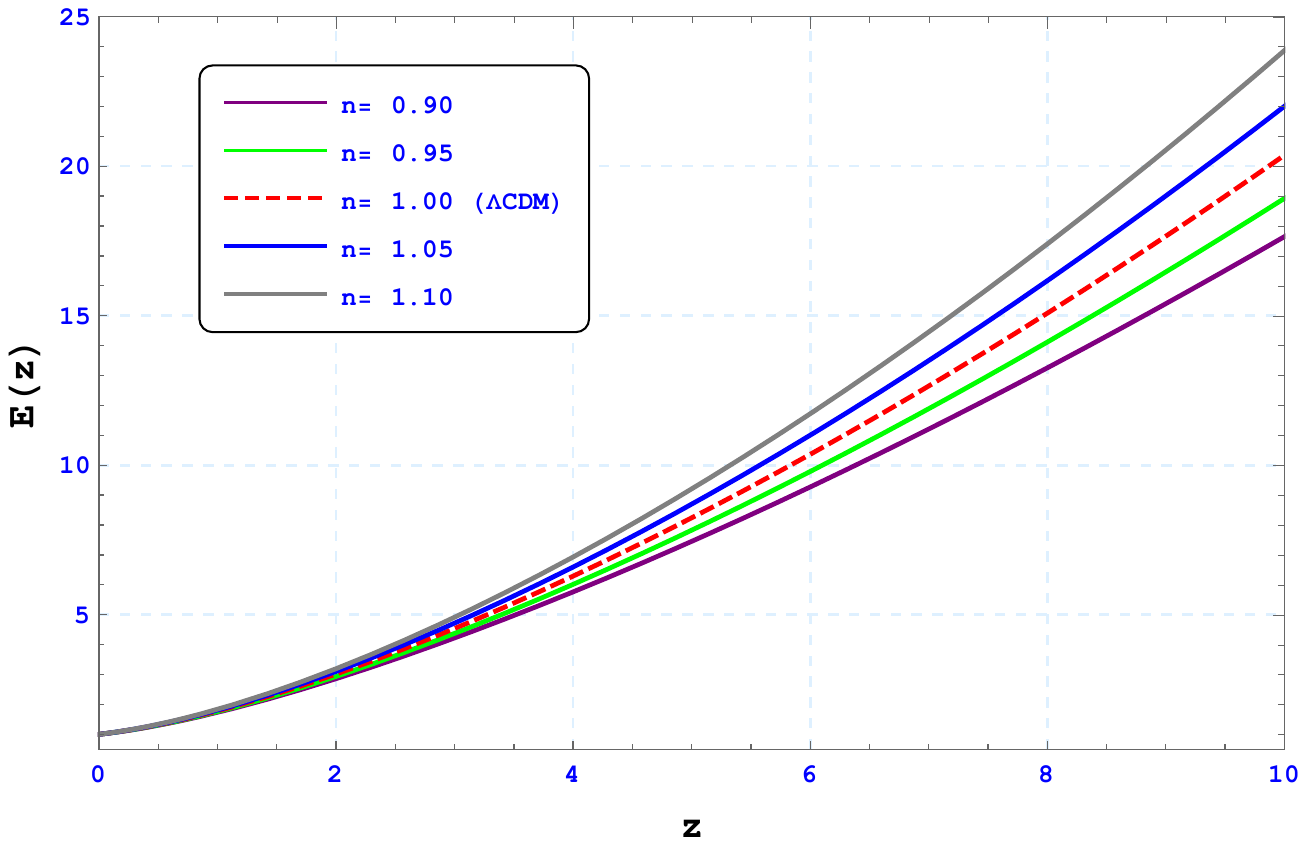}
    \caption{\small(Color online) Plot for normalized Hubble parameter $E(z)$ versus redshift $z$ with different model parameter $n$}
    \label{Ez.pdf}
\end{figure}
We assume the pressureless matter and DE are inherently non-interacting in nature; thereby, they follow separate conservation equations given by
\begin{subequations}\label{contituity_eqn_seperate}
    \begin{align}
       \dot{\rho}_m + 3\frac{\dot{a}}{a}\rho_m &= 0 \,,\\
       \dot{\rho}_\Lambda &= 0 \,,  
    \end{align}
\end{subequations}
respectively, as derived from Eq. (\ref{continuity_equation}). The following solutions for Eq. (\ref{contituity_eqn_seperate}) can be obtained as
\begin{subequations}\label{solutions_matter_DE}
    \begin{align}
       \rho_m &= \rho_{m,0}a^{-3} =\rho_{m,0}(1+z)^{3} \,,\\
       \rho_\Lambda &= \rho_{\Lambda,0} \,.  
    \end{align}
\end{subequations}
Here, $\rho_{m,0}$ and $\rho_{\Lambda,0}$ are the present-day ($a=1$) densities of matter and DE, respectively, and $z = \frac{1}{a}-1$ is the redshift. The scale factor for today is considered $a_0 = 1$. Substituting (\ref{solutions_matter_DE}) in Eq. (\ref{mod_Friedmann_eqn}) for a flat Universe scenario, i.e., $\kappa = 0$, we obtain the Hubble parameter $H(z)$ and normalized Hubble parameter $E(z)$ in the GMHE-inspired modified cosmology as follows
\begin{subequations}
    \begin{align}
       H(z) &= \Gamma_{\gamma,n}^\frac{1}{3-n}[\rho_{m} + \rho_{\Lambda}]^\frac{1}{3-n} = \Gamma_{\gamma,n}^\frac{1}{3-n}[\rho_{m,0}(1+z)^{3} + \rho_{\Lambda,0}]^\frac{1}{3-n} \,,\label{Hubble}\\
       E(z) &\overset{\text{def}}{=}  \frac{H(z)}{H_0} =  [\Omega_{m,0}(1+z)^{3}+\Omega_{\Lambda,0}]^\frac{1}{3-n} \label{Normalized Hubble}\,,  
    \end{align}
\end{subequations}
where $H_0 = 100\,\mathrm{h} \text{ km s}^{-1} \text{ Mpc}^{-1}$ with $\mathrm{h} = 0.6766 \pm 0.0042$, $\Omega_{m,0} = \frac{\rho_m}{\rho_{cr}} = 0.3111 \pm 0.0056$ and $\Omega_{\Lambda,0} = \frac{\rho_\Lambda}{\rho_{cr}} = 0.6889 \pm 0.0056$ \cite{aghanim2020planck,aghanim2021erratum} are the present-day ($z=0$) Hubble parameter, density parameters of matter and DE, respectively, and $\rho_{cr}$ is the critical density in standard $\Lambda$CDM cosmology. At the present Universe, $E(z=0) = 1$. In Fig. \ref{Ez.pdf}, the redshift evolution of the normalized Hubble parameter is shown for different values of model parameter $n$. In the GMHE-inspired modified cosmology, the normalized Hubble parameter shows an incline as the model parameter $n$ shifts from lower to higher values. Moreover, for $n<1$ ($n>1$), the normalized Hubble parameter displays a gentler (sharper) slope, showing that the expansion rate of this model Universe becomes slower (faster) in contrast to the standard $\Lambda$CDM framework.
%%%%%%%%%%%%%%%%%%%%%%%%%%%%%%%%%%%%%%%%%%%%%%%%%%%%%%%%%%%%%%%%%%%%%%%%%%%%%%%%%%%%%%%%%%%%%%%%%%%%%%%%%%%%%%%%%%%%%%%%%%%%%%%%%%%%%%%%%%%%%%
\begin{figure}[htbp!]
    \centering
    \includegraphics[width=10cm,height=8cm]{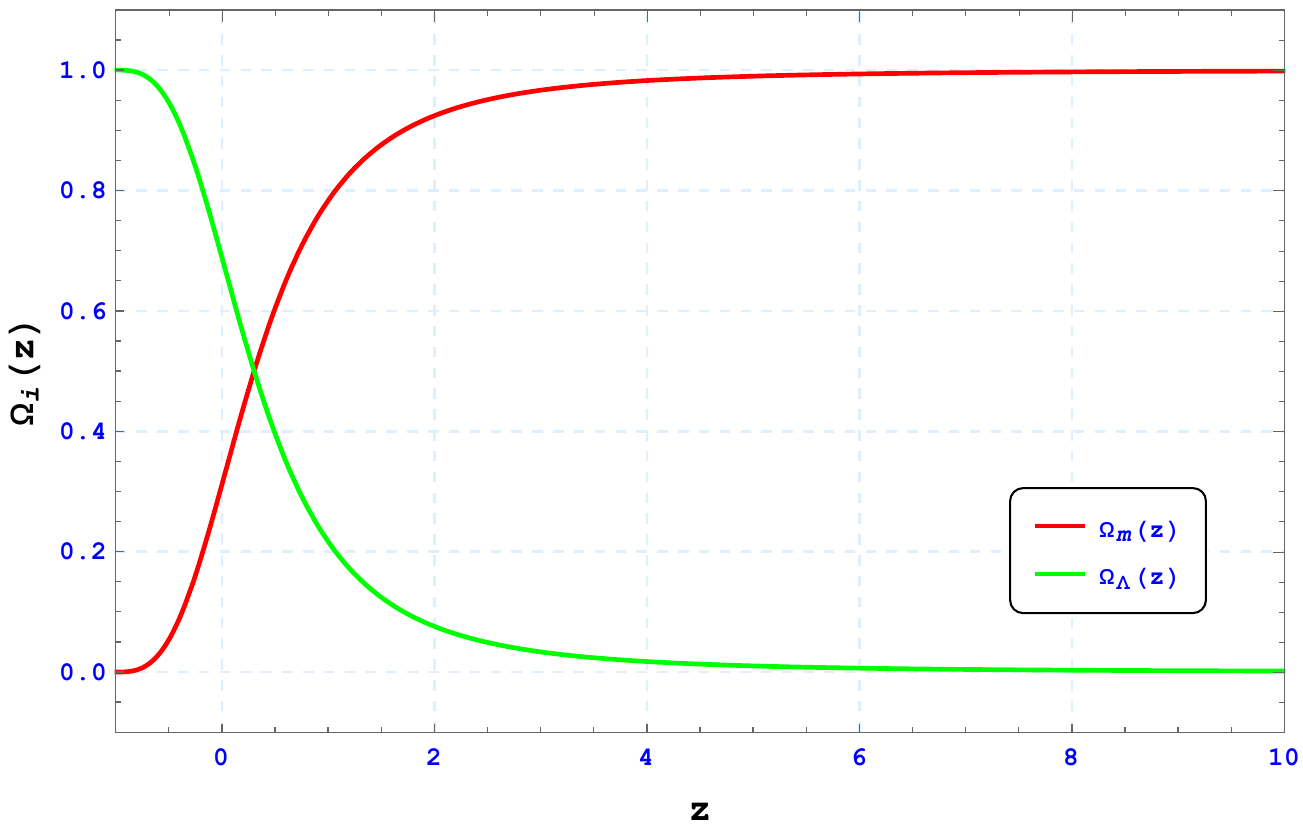}
    \caption{\small(Color online) Plot for cosmological density parameters $\Omega_i(z)\, (i\equiv m, \Lambda)$ versus redshift $z$}
    \label{OMEGAiz.pdf}
\end{figure}
%%%%%%%%%%%%%%%%%%%%%%%%%%%%%%%%%%%%%%%%%%%%%%%%%%%%%%%%%%%%%%%%%%%%%%%%%%%%%%%%%%%%%%%%%%%%%%%%%%%%%%%%%%%%%%%%%%%%%%%%%%%%%%%%%%%%%%%%%%%%%%
\par From Eqs. (\ref{density_param}) and (\ref{Hubble}), the evolution of matter and DE density parameters $\Omega^{mod}_m(z)$ and $\Omega^{mod}_{\Lambda}(z)$ in the GMHE-inspired modified cosmology can then be expressed as
\begin{subequations}\label{density_parameters}
    \begin{align}
       \Omega^{mod}_m(z) &= \frac{\rho_m}{\rho_m + \rho_\Lambda} = \frac{\Omega_{m,0}(1+z)^{3}}{\Omega_{m,0}(1+z)^{3}+\Omega_{\Lambda,0}} = \Omega_m(z) \,,\label{matter_density_parameter}\\
       \Omega^{mod}_{\Lambda}(z) &= \frac{\rho_\Lambda}{\rho_m + \rho_\Lambda} = \frac{\Omega_{\Lambda,0}}{\Omega_{m,0}(1+z)^{3}+\Omega_{\Lambda,0}} = \Omega_{\Lambda}(z)\,\label{DE_density_parameter},  
    \end{align}
\end{subequations}
respectively, which are independent of any model parameters. They resemble the standard flat $\Lambda$CDM cosmology and follow the relation $\Omega_m + \Omega_\Lambda = 1$. The cosmological density parameters in Eq. (\ref{density_parameters}) have been illustrated in Fig. \ref{OMEGAiz.pdf}, and at a redshift $z_{eq} = 0.32635$, the matter-DE equality epoch is indicated, just like in the $\Lambda$CDM model. We observe $\Omega_\Lambda \to 1$ and $\Omega_m \to 0$ as $z\to -1$, indicating a scenario in which DE entirely dominates the energy budget of the Universe in the distant future, although the early Universe was under matter domination.
%%%%%%%%%%%%%%%%%%%%%%%%%%%%%%%%%%%%%%%%%%%%%%%%%%%%%%%%%%%%%%%%%%%%%%%%%%%%%%%%%%%%%%%%%%%%%%%%%%%%%%%%%%%%%%%%%%%%%%%%%%%%%%%%%%%%%%%%%%%%%%
\begin{figure}[htbp!]
    \centering
    \includegraphics[width=10cm,height=8cm]{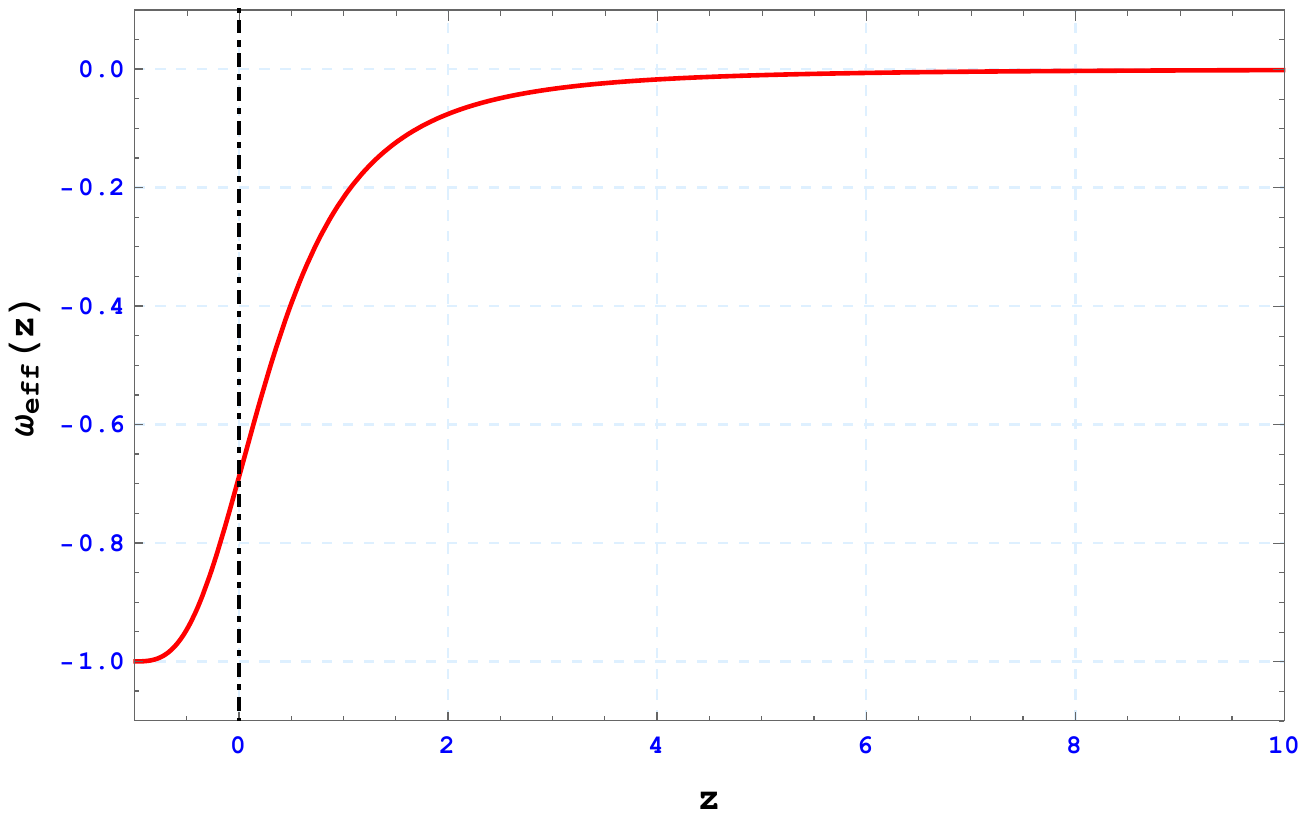}
    \caption{\small(Color online) Plot for effective equation of state parameter $w_{eff}$ versus redshift $z$}
    \label{omegaeffz.pdf}
\end{figure}
%%%%%%%%%%%%%%%%%%%%%%%%%%%%%%%%%%%%%%%%%%%%%%%%%%%%%%%%%%%%%%%%%%%%%%%%%%%%%%%%%%%%%%%%%%%%%%%%%%%%%%%%%%%%%%%%%%%%%%%%%%%%%%%%%%%%%%%%%%%%%%
\par The nature of the effective equation of state parameter summarizes the collective behavior of the cosmic components driving the expansion. Its evolution provides a simple and intuitive way to track deviations from matter domination and the emergence of cosmic acceleration. Considering the mixture of pressureless matter and DE as a single fluid, the effective equation of state parameter $\omega_{\, eff}(z)$ can be expressed in GMHE-inspired modified cosmology as
\begin{equation}
    \omega_{\,eff}(z) \overset{\text{def}}{=} \frac{P_m + P_\Lambda}{c^2(\rho_m + \rho_\Lambda)} = -\frac{\Omega_{\Lambda,0}}{\Omega_{m,0}(1+z)^3 + \Omega_{\Lambda,0}}\, ,
\end{equation}
which is independent of our model parameters and exactly matches the flat $\Lambda$CDM model. The evolution of the effective equation of state with redshift is shown in Fig. \ref{omegaeffz.pdf}. We see, as $z\to -1$, $\omega_{\, eff} = -1$, which signifies a shift to the de-Sitter phase in the distant future. At present, where $z\to 0$, we  get $\omega_{\, eff} =\Omega_{\Lambda,0} = -0.6889$, whilst in the very early Universe $z\to \infty$, we have $\omega_{\, eff} = 0$. This suggests that the Universe experiences deceleration at an early stage, while experiencing acceleration at a later stage.
%%%%%%%%%%%%%%%%%%%%%%%%%%%%%%%%%%%%%%%%%%%%%%%%%%%%%%%%%%%%%%%%%%%%%%%%%%%%%%%%%%%%%%%%%%%%%%%%%%%%%%%%%%%%%%%%%%%%%%%%%%%%%%%%%%%%%%%%%
\par Cosmography \cite{visser2005cosmography,visser2004jerk} is a popular geometric-based method to test the divergence of various modified cosmological models from the standard $\Lambda$CDM framework. The scale factor for the FLRW Universe can be expanded around a current epoch $t_0$ in a Taylor series as $a(t)=\sum_{i=0}^{\infty}\frac{a^{(i)}(t_0)}{i!}(t-t_0)^i$, where $a^{(i)}(t_0)$ refers to the $i^{\text{th}}$ derivative of scale factor $a(t)$ at $t=t_0$. The coefficients of the Taylor expansion are related to the geometric quantities $H = \frac{\dot{a}}{a}$, $q = -\frac{\overset{\cdot\cdot}{a}}{aH^2}$, $j = \frac{\overset{\cdot\cdot\cdot}{a}}{aH^3}$, $s = \frac{\overset{\cdot\cdot\cdot\cdot}{a}}{aH^4}$, $\ell = \frac{\overset{\cdot\cdot\cdot\cdot\cdot}{a}}{aH^5}$, $m = \frac{\overset{\cdot\cdot\cdot\cdot\cdot\cdot}{a}}{aH^6}$, and e.t.c. are so-called the \textit{cosmographic parameters}. Here, the newly introduced quantities $q$, $j$, $s$, $\ell$, and $m$ are referred to as the \textit{deceleration}, \textit{jerk}, \textit{snap}, \textit{lerk}, and \textit{$m$}-parameters \cite{pan2018astronomical}. Prior to the discovery of cosmic acceleration, $H$ was the primary focus of the observation. As it evolves with time, the next higher order derivative of the scale factor, i.e., $q$, accounts for its evolution. Now we observe $q$ itself evolves, then $j$ emerges as the automatic choice for the researchers to study \cite{mukherjee2016parametric}. This sequence can also be further extended to a few more kinematic quantities.
\par
\begin{figure}[htbp!]
    \centering
    \includegraphics[width=10cm,height=8cm]{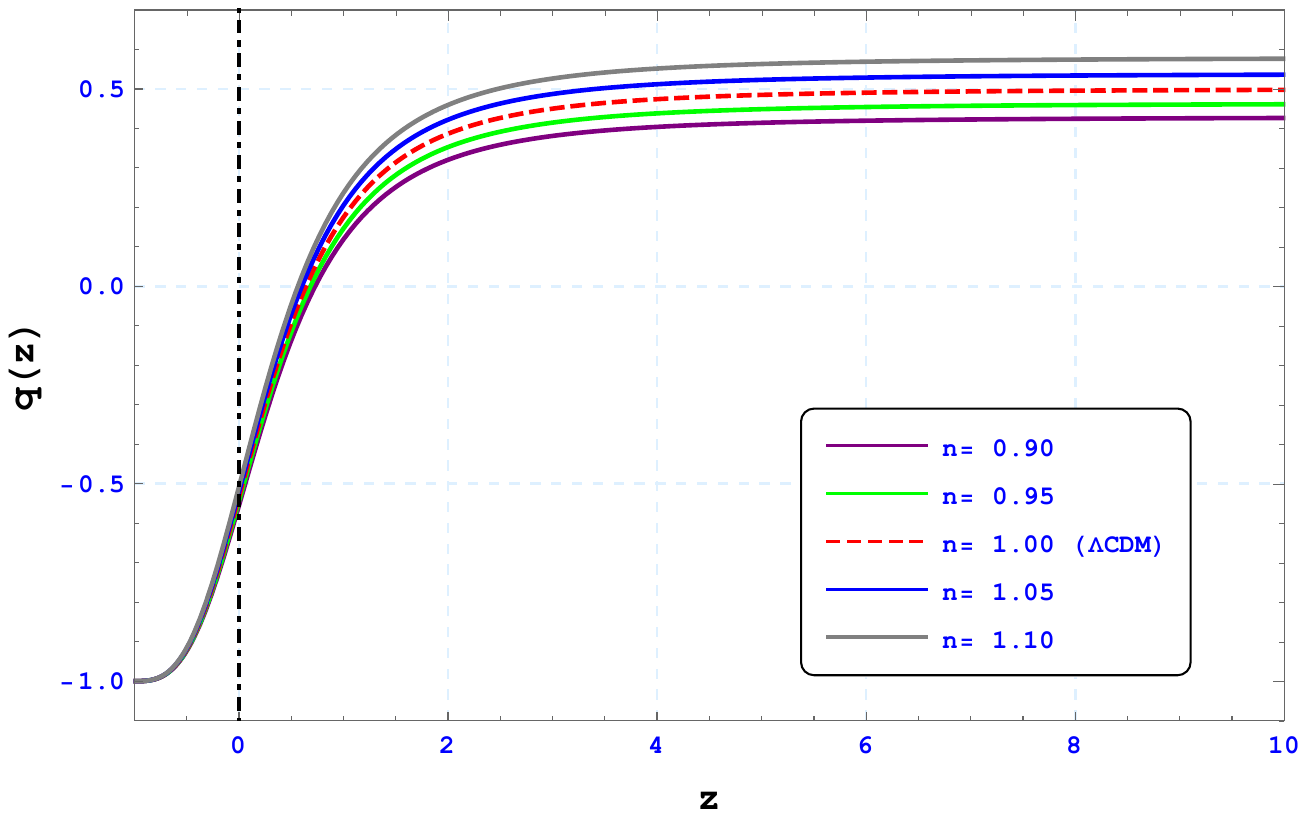}
    \caption{\small(Color online) Plot for deceleration paramater $q$ versus redshift $z$ with different model parameter $n$}
    \label{qz.pdf}
\end{figure}

 The deceleration parameter $q$ provides a kinematical measure of the cosmic expansion by quantifying whether the Universe undergoes accelerated or decelerated expansion. Its evolution directly traces the transition between these regimes and reflects the relative influence of the dominant energy components driving the expansion. As a result, it provides a fundamental way to probe DE dynamics/modification in cosmology and departures from the standard cosmological picture. The deceleration parameter $q(z)$ for the GMHE-inspired modified cosmology can be expressed in relation to redshift as follows
\begin{align}\label{deceleration_param}
       q(z) &\overset{\text{def}}{=} -\frac{\overset{\cdot\cdot}{a}}{a H^2} = -\left(1+\frac{\dot{H}}{H}\right) = -1 + \frac{(1+z)}{E(z)}\frac{dE(z)}{dz}= -1 + \left(\frac{3}{3-n}\right)\frac{\Omega_{m,0}(1+z)^{3}}{\Omega_{m,0}(1+z)^{3}+\Omega_{\Lambda,0}} \, . 
\end{align}
The transition redshift $z_{\mathrm{tr}}$ refers to the epoch in cosmic history when the expansion rate switches from deceleration ($q>0$) to acceleration ($q<0$), delineated by the condition $q(z=z_{\, tr})=0$. This yields
\begin{equation}
z_{\mathrm{tr}} =\left[\left(\frac{3-n}{n}\right)\,
\frac{\Omega_{\Lambda,0}}{\Omega_{m,0}}\right]^{\tfrac{1}{3}}
-1 \,.
\end{equation}
The Fig. \ref{qz.pdf} effectively showcases the transition from deceleration phase to acceleration phase for different values of $n$ at a redshift around $z_{\mathrm{tr}}\approx 0.64$, which is compatible with observations \cite{myrzakulov2024linear,roman2019constraints,al2018observational}. It is also seen that $z_{\mathrm{tr}}$ depends on model parameter $n$. In particular, we find that $z_{\mathrm{tr}}$ decreases (increases) for $n > 1$ ($n < 1$), indicating that the starting point of cosmic acceleration occurs later (earlier) in GMHE-inspired modified cosmology relative to the standard $\Lambda$CDM profile. We also observe that the deceleration parameter increases with the increment of $n$ at the early epoch $n$. At $z \to -1$, the modified model converges to the $\Lambda$CDM model. As we observe $q \to -1$ when $z \to -1$, therefore, the criterion for the Universe to approach thermodynamic equilibrium in the distant future is not violated \cite{del2012three}.
%%%%%%%%%%%%%%%%%%%%%%%%%%%%%%%%%%%%%%%%%%%%%%%%%%%%%%%%%%%%%%%%%%%%%%%%%%%%%%%%%%%%%%%%%%%%%%%%%%%%%%%%%%%%%%%%%%%%%%%%%%%%%%%%%%%%%%%%%%%%%%
%%%%%%%%%%%%%%%%%%%%%%%%%%%%%%%%%%%%%%%%%%%%%%%%%%%%%%%%%%%%%%%%%%%%%%%%%%%%%%%%%%%%%%%%%%%%%%%%%%%%%%%%%%%%%%%%%%%%%%%%%%%%%%%%%%%%%%%%%
\par
\begin{figure}[htbp!]
    \centering
    \includegraphics[width=10cm,height=8cm]{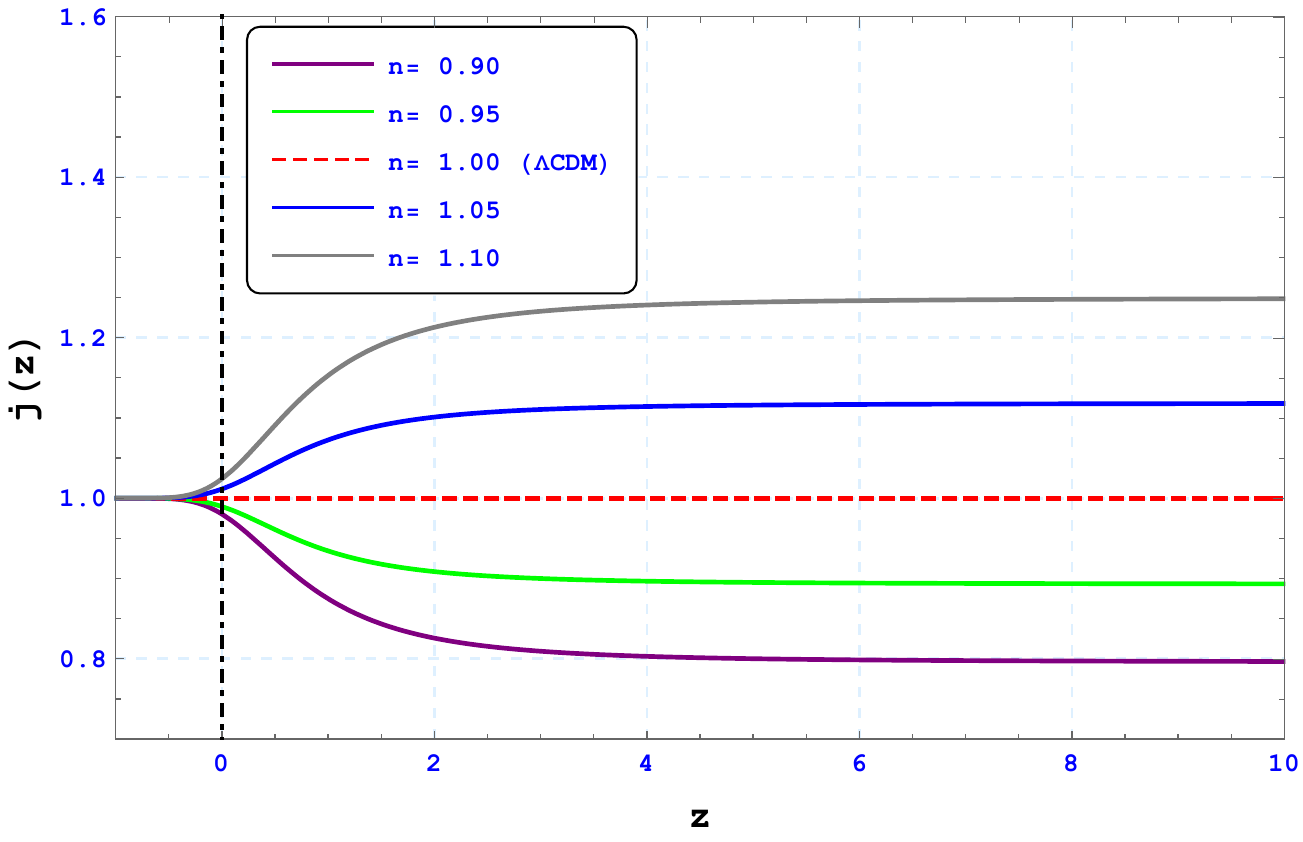}
    \caption{\small(Color online) Plot for jerk parameter $j(z)$ versus redshift $z$ with different model parameter $n$}
    \label{j(z).pdf}
\end{figure}

 The jerk parameter offers a higher-order kinematical perspective on the cosmic expansion, complementing the information provided by the Hubble and deceleration parameters. Although the deceleration-acceleration phase transition is dictated by the second derivative of the scale factor (deceleration parameter), the jerk parameter captures how the acceleration itself evolves during this change. In terms of redshift, the jerk parameter $j(z)$ in the GMHE-inspired modified cosmology can be written as
\begin{align}\label{jerk_parameter}
j(z) 
&\overset{\text{def}}{=} 
   \frac{\overset{\cdot\cdot\cdot}{a}}{a H^3} 
   = q(z)\,[2q(z)+1] + (1+z)\frac{dq(z)}{dz}
  \nonumber \\[6pt]
&= 
\frac{1}{
   \left[\Omega_{m,0}(1+z)^3 + \Omega_{\Lambda,0}\right]^2
   }
\Bigg[
      \frac{n(3+n)}{(3-n)^2}\,
      \Omega_{m,0}^2 (1+z)^6
      + 2\,\Omega_{m,0}\,\Omega_{\Lambda,0} (1+z)^3 + \Omega_{\Lambda,0}^2
\Bigg]\,.
\end{align}
 At the transition epoch, the relation
\begin{equation}
j(z_{\mathrm{tr}}) = (1+z_{\mathrm{tr}})\frac{dq}{dz}\Bigg\rfloor_{z=z_{\mathrm{tr}}}
\end{equation}
shows that the jerk parameter effectively measures the smoothness of the transition from decelerated to accelerated expansion. In Fig. \ref{j(z).pdf}, we have plotted the redshift evolution of the jerk parameter for different values of the model parameter $n$. Within the standard $\Lambda$CDM framework, the jerk remains constant with $j(z)\big\rfloor_{n = 1}=1$ at all redshifts. Any deviation from this value, therefore, provides a clear indication of physics beyond the standard cosmological model. We observe that the jerk parameter increases with $n$, and at $z \to -1$ the modified model coincides with the $\Lambda$CDM model. It is also a notable fact that the jerk parameter is always positive for any redshift and $n$, suggesting the accelerated expansion of the Universe. Interestingly, the jerk parameter deviates from unity at $z=0$, indicating a clear departure from $\Lambda$CDM predictions at the current epoch. This provides an important characteristic of GMHE-inspired modified cosmology, that the modification to SMHR may influence late-time acceleration.
%%%%%%%%%%%%%%%%%%%%%%%%%%%%%%%%%%%%%%%%%%%%%%%%%%%%%%%%%%%%%%%%%%%%%%%%%%%%%%%%%%%%%%%%%%%%%%%%%%%%%%%%%%%%%%%%%%%%%%%%%%%%%%%%%%%%%%%
%%%%%%%%%%%%%%%%%%%%%%%%%%%%%%%%%%%%%%%%%%%%%%%%%%%%%%%%%%%%%%%%%%%%%%%%%%%%%%%%%%%%%%%%%%%%%%%%%%%%%%%%%%%%%%%%%%%%%%%%%%%%%%%%%%%%%%%
\par
\begin{figure}[htbp!]
    \centering
    \includegraphics[width=10cm,height=8cm]{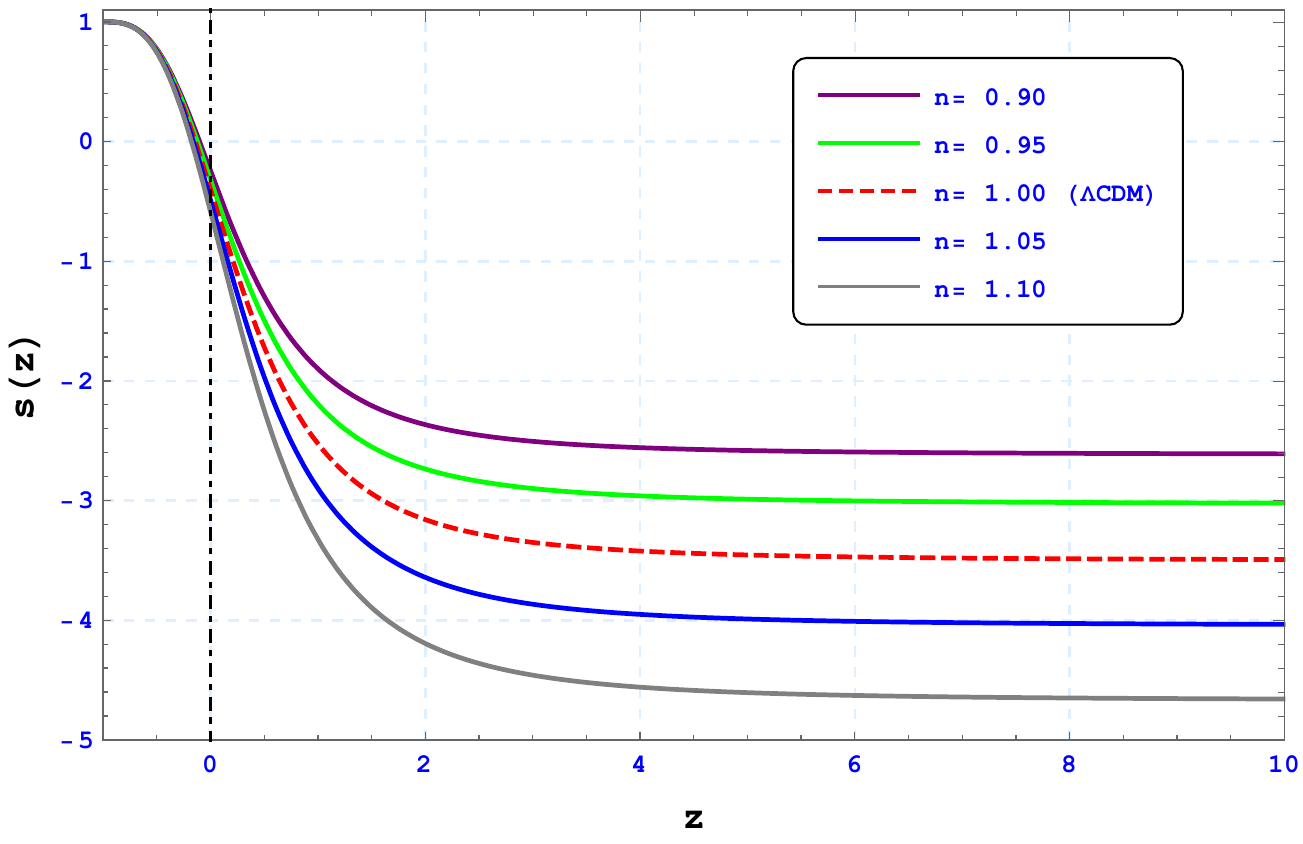}
    \caption{\small(Color online) Plot for snap parameter $s(z)$ versus redshift $z$ with different model parameter $n$}
    \label{s(z).pdf}
\end{figure}

 Broadening the information conveyed by the Hubble, deceleration, and jerk parameters, the next higher-order cosmographic parameter, i.e., snap parameter, also offers a higher-order kinematical illustration of the cosmological expansion. While the deceleration parameter indicates the deceleration-acceleration changeover of our Universe, and the jerk describes the evolutionary behavior of the acceleration, the snap parameter characterizes the evolution of the jerk parameter. Hence, the snap parameter offers another insightful probe of divergence from the standard $\Lambda$CDM profile. Regarding redshift, the snap parameter $s$ in the GMHE-inspired modified cosmology can be expressed as
\begin{align}\label{snap_parameter}
s(z)
&\overset{\text{def}}{=} \frac{\ddddot a}{aH^4}
= -j(z)\,[3q(z)+2] - (1+z)\frac{dj(z)}{dz}
\nonumber \\[6pt]
&= -\frac{1}{
\left[\Omega_{m,0}(1+z)^3+\Omega_{\Lambda,0}\right]^3
}
\Bigg[
\frac{n(3+n)(6+n)}{(3-n)^3}\,
\Omega_{m,0}^3(1+z)^9
- \frac{3(n^2-15n+6)}{(3-n)^2}\,
\Omega_{m,0}^2\Omega_{\Lambda,0}(1+z)^6
\nonumber\\[4pt]
&\qquad\quad
+ \frac{3n}{(3-n)}\,
\Omega_{m,0}\Omega_{\Lambda,0}^2(1+z)^3
- \Omega_{\Lambda,0}^3
\Bigg]\,.
\end{align}
In Fig. \ref{s(z).pdf}, we have plotted the redshift evolution of the snap parameter for different values of the model parameter $n$. We notice that the snap parameter shows a decrease with an increment of $n$. For the flat $\Lambda$CDM profile, the snap parameter rises from \(s=-3.5\) during the matter-domination, passes through \(s=-2\) at the deceleration to acceleration transition epoch, and approaches the asymptotic de-Sitter value \(s=1\), highlighting the divergence between the $\Lambda$CDM and the GMHE-inspired modified gravity scenarios apart from the distant future ($z \to -1$).
%%%%%%%%%%%%%%%%%%%%%%%%%%%%%%%%%%%%%%%%%%%%%%%%%%%%%%%%%%%%%%%%%%%%%%%%%%%%%%%%%%%%%%%%%%%%%%%%%%
%%%%%%%%%%%%%%%%%%%%%%%%%%%%%%%%%%%%%%%%%%%%%%%%%%%%%%%%%%%%%%%%%%%%%%%%%%%%%%%%%%%%%%%%%%%%%%%%%%%%%%%%%%%%%%%%%%%%%%%%%%%%%%%%%%%%%%%%%
\par
\begin{figure}[htbp!]
    \centering
    \includegraphics[width=10cm,height=8cm]{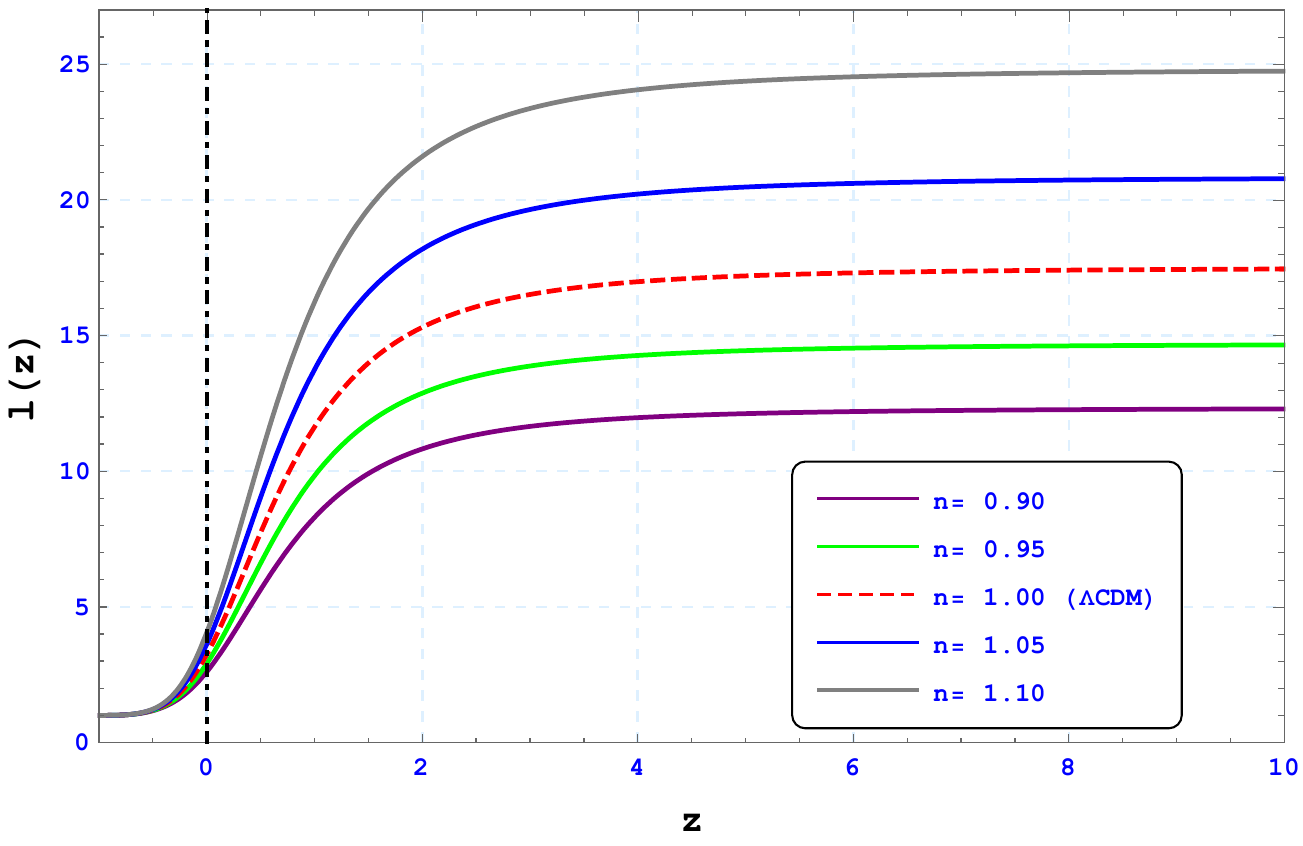}
    \caption{\small(Color online) Plot for lerk parameter $l(z)$ versus redshift $z$ with different model parameter $n$}
    \label{l(z).pdf}
\end{figure}

 In a similar manner, we introduce the next higher-order cosmographic parameter, the lerk parameter, in order to measure the rate of change of snap parameter, offering a subtle probe to detect tiny deviations from the $\Lambda$CDM scenario. In relation to redshift, the lerk parameter $\ell$ in GMHE-inspired modified cosmology reads
\begin{align}\label{lerk_parameter}
\ell(z)
&\overset{\text{def}}{=} \frac{\overset{\cdot\cdot\cdot\cdot\cdot}{a}}{aH^5}
= -s(z)\,[4q(z)+3] - (1+z)\frac{ds(z)}{dz}
\nonumber \\[6pt]
&=
\frac{1}{
\big[\Omega_{m,0}(1+z)^3 + \Omega_{\Lambda,0}\big]^4
}
\Bigg[
\frac{n(3+n)(6+n)(9+n)}{(3-n)^4}\,
\Omega_{m,0}^4(1+z)^{12}
- \frac{4n\big(n^2-63n-18\big)}{(3-n)^3}\,
\Omega_{\Lambda,0}\Omega_{m,0}^3(1+z)^9
\nonumber\\[4pt]
&\qquad\quad
+ \frac{6\big(n^2+33n-15\big)}{(3-n)^2}\,
\Omega_{\Lambda,0}^2\Omega_{m,0}^2(1+z)^6
+ \frac{2(9-2n)}{(3-n)}\,
\Omega_{\Lambda,0}^3\Omega_{m,0}(1+z)^3
+ \Omega_{\Lambda,0}^4
\Bigg]\,.
\end{align}
We have depicted the redshift evolution of the lerk parameter for different values of the model parameter $n$ in Fig. \ref{l(z).pdf}. We observe that the higher the value of $n$, the higher the lerk parameter across all redshifts. Like the jerk parameter, the lerk parameter is always positive for any redshift and $n$, indicating the accelerated expansion of the Universe. The modified cosmological model coincides with the standard $\Lambda$CDM model at $z \to -1$.
%%%%%%%%%%%%%%%%%%%%%%%%%%%%%%%%%%%%%%%%%%%%%%%%%%%%%%%%%%%%%%%%%%%%%%%%%%%%%%%%%%%%%%%%%%%%%%%%%%%%%%%%%%%%%%%%%%%%%%%%%%%%%%%%%%%%%%%%%
%%%%%%%%%%%%%%%%%%%%%%%%%%%%%%%%%%%%%%%%%%%%%%%%%%%%%%%%%%%%%%%%%%%%%%%%%%%%%%%%%%%%%%%%%%%%%%%%%%%%%%%%%%%%%%%%%%%%%%%%%%%%%%%%%%%%%%%%%
\par
\begin{figure}[htbp!]
    \centering
    \includegraphics[width=10cm,height=8cm]{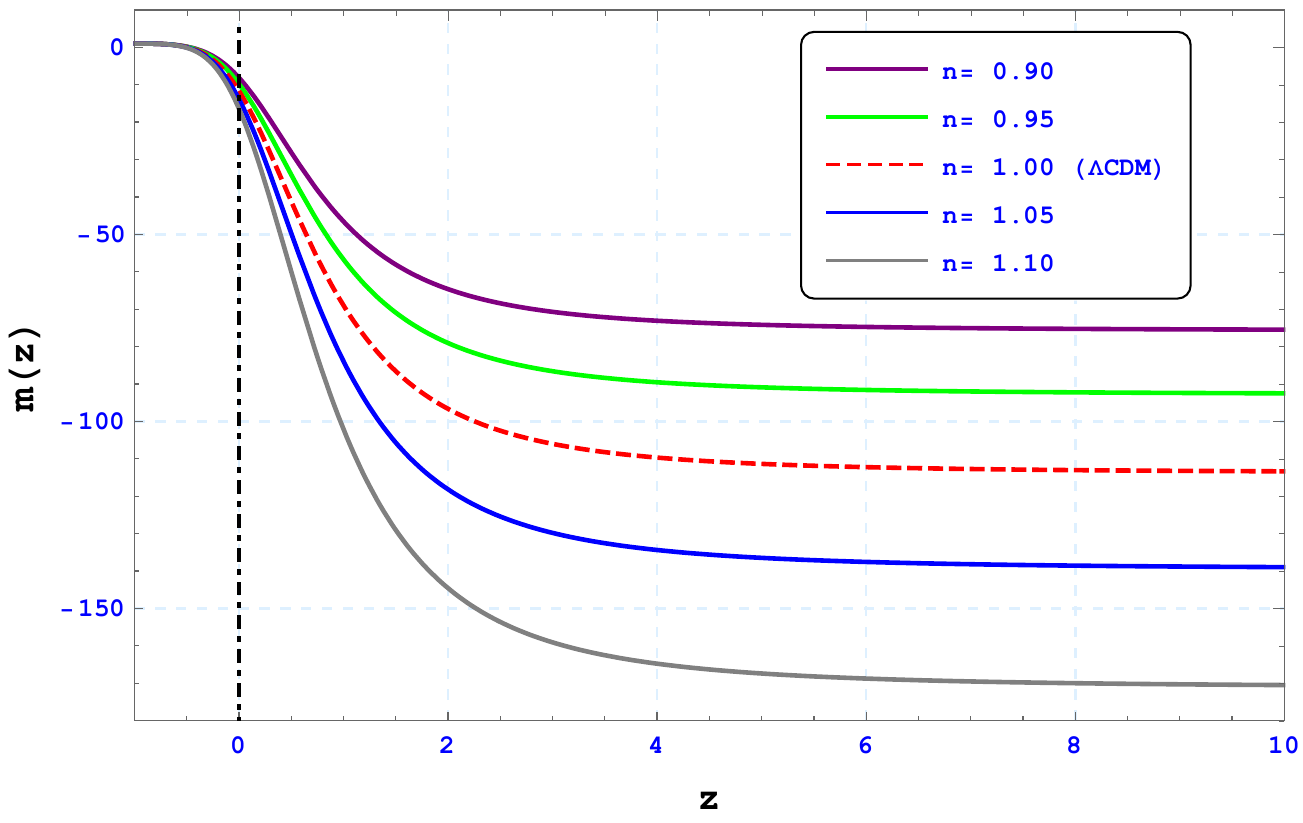}
    \caption{\small(Color online) Plot for m--parameter $l(z)$ versus redshift $z$ with different model parameter $n$}
    \label{m(z).pdf}
\end{figure}

 One can extend this chain of derivatives to find the next higher-order kinematical parameters. However, we restrict ourselves to the sixth derivative of the scale factor, called the $m$-parameter, which records the evolution of the lerk parameter and is also a very sensitive probe of deviations from standard cosmology. The $m$-parameter in the GMHE-inspired modified cosmology can be expressed in terms of redshift as
\begin{align}\label{m_parameter}
m(z)
&\overset{\text{def}}{=} \frac{\overset{\cdot\cdot\cdot\cdot\cdot\cdot}{a}}{aH^6}
= -\ell(z)\,[5q(z)+4] - (1+z)\frac{d\ell(z)}{dz}
\nonumber \\[6pt]
&=
\frac{1}{
\big[\Omega_{m,0}(1+z)^3 + \Omega_{\Lambda,0}\big]^5
}
\Bigg[
-\frac{n(3+n)(6+n)(9+n)(12+n)}{(3-n)^5}\,
\Omega_{m,0}^5(1+z)^{15}
\nonumber\\[4pt]
&\qquad\quad
+ \frac{n(5n^3-1194n^2-2133n-1998)}{(3-n)^4}\,
\Omega_{m,0}^4\Omega_{\Lambda,0}(1+z)^{12}
\nonumber\\[4pt]
&\qquad\quad
- \frac{2(5n^3+1683n^2-621n+270)}{(3-n)^3}\,
\Omega_{m,0}^3\Omega_{\Lambda,0}^2(1+z)^9
\nonumber\\[4pt]
&\qquad\quad
+ \frac{10(n^2-123n+72)}{(3-n)^2}\,
\Omega_{m,0}^2\Omega_{\Lambda,0}^3(1+z)^6
- \frac{(12+5n)}{(3-n)}\,
\Omega_{m,0}\Omega_{\Lambda,0}^4(1+z)^3
+ \Omega_{\Lambda,0}^5
\Bigg]\,.
\end{align}
The evolution of the $m$ parameter for different values of the model parameter $n$ is displayed in Fig. \ref{m(z).pdf}. We see that the $m$ parameter decreases as the model parameter $n$ increases, and it is always negative. At $z \to -1$, the modified model aligns with the $\Lambda$CDM model.
%%%%%%%%%%%%%%%%%%%%%%%%%%%%%%%%%%%%%%%%%%%%%%%%%%%%%%%%%%%%%%%%%%%%%%%%%%%%%%%%%%%%%%%%%%%%%%%%%%%%%%%%%%%%%%%%%%%%%%%%%%%%%%%%%%%%%%%%%
%%%%%%%%%%%%%%%%%%%%%%%%%%%%%%%%%%%%%%%%%%%%%%%%%%%%%%%%%%%%%%%%%%%%%%%%%%%%%%%%%%%%%%%%%%%%%%%%%%%%%%%%%%%%%%%%%%%%%%%%%%%%%%%%%%%%%%%%%
\begin{figure}[htbp!]
    \centering
    \includegraphics[width=10cm,height=8cm]{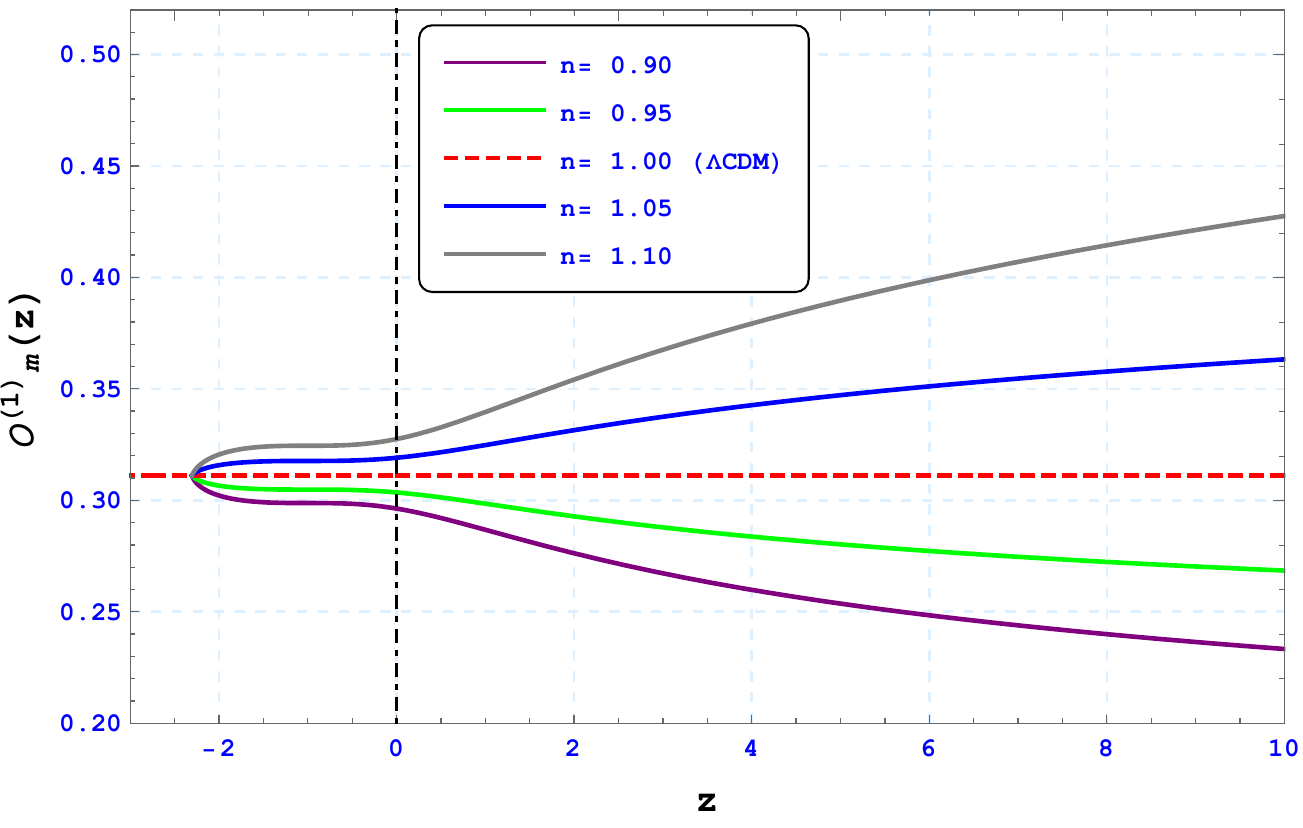}
    \caption{\small(Color online) Plot for $\mathcal{O}^{(1)}_m(z)$ parameter versus redshift $z$ with different model parameter $n$}
    \label{O1m(z).pdf}
\end{figure}
%%%%%%%%%%%%%%%%%%%%%%%%%%%%%%%%%%%%%%%%%%%%%%%%%%%%%%%%%%%%%%%%%%%%%%%%%%%%%%%%%%%%%%%%%%%%%%%%%%%%%%%%%%%%%%%%%%%%%%%%%%%%%%%%%%%%%%%%%
\begin{figure}[htbp!]
    \centering
    \includegraphics[width=10cm,height=8cm]{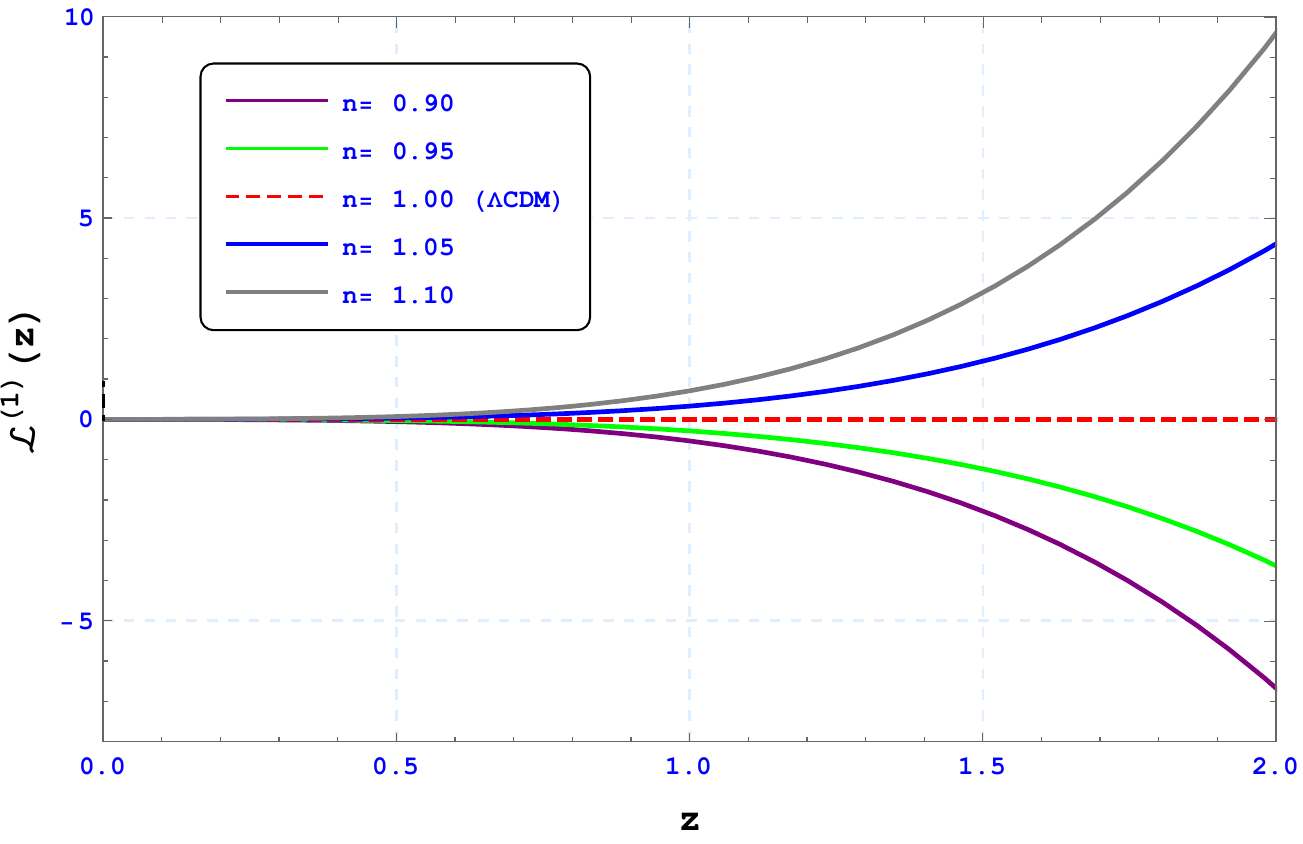}
    \caption{\small(Color online) Plot for $\mathcal{L}^{(1)}(z)$ parameter versus redshift $z$ with different model parameter $n$}
    \label{L1(z).pdf}
\end{figure}
%%%%%%%%%%%%%%%%%%%%%%%%%%%%%%%%%%%%%%%%%%%%%%%%%%%%%%%%%%%%%%%%%%%%%%%%%%%%%%%%%%%%%%%%%%%%%%%%%%%%%%%%%%%%%%%%%%%%%%%%%%%%%%%%%%%%%%%%%
\begin{figure}[htbp!]
    \centering
    \includegraphics[width=10cm,height=8cm]{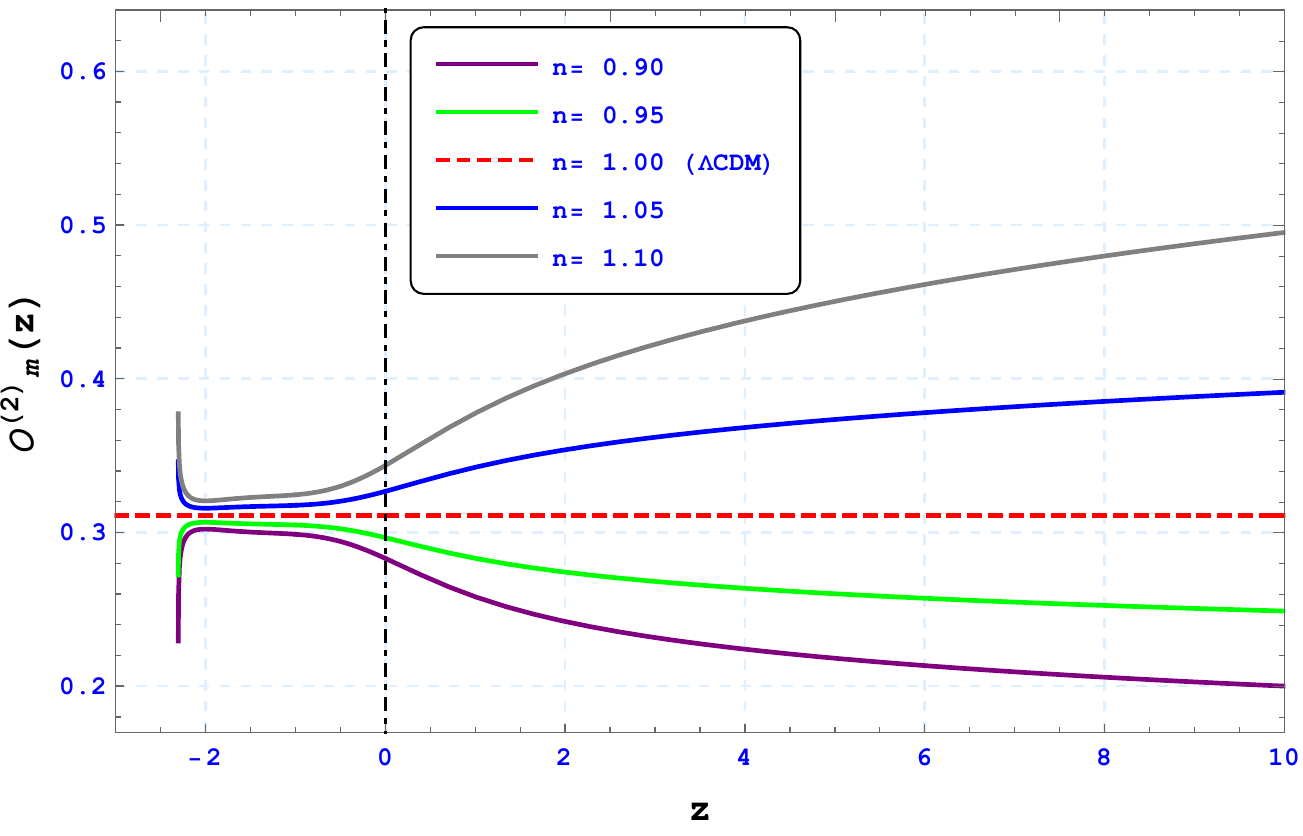}
    \caption{\small(Color online) Plot for $\mathcal{O}^{(2)}_m(z)$ parameter versus redshift $z$ with different model parameter $n$}
    \label{O2m(z).pdf}
\end{figure}
%%%%%%%%%%%%%%%%%%%%%%%%%%%%%%%%%%%%%%%%%%%%%%%%%%%%%%%%%%%%%%%%%%%%%%%%%%%%%%%%%%%%%%%%%%%%%%%%%%%%%%%%%%%%%%%%%%%%%%%%%%%%%%%%%%%%%%%%%
\begin{figure}[htbp!]
    \centering
    \includegraphics[width=10cm,height=8cm]{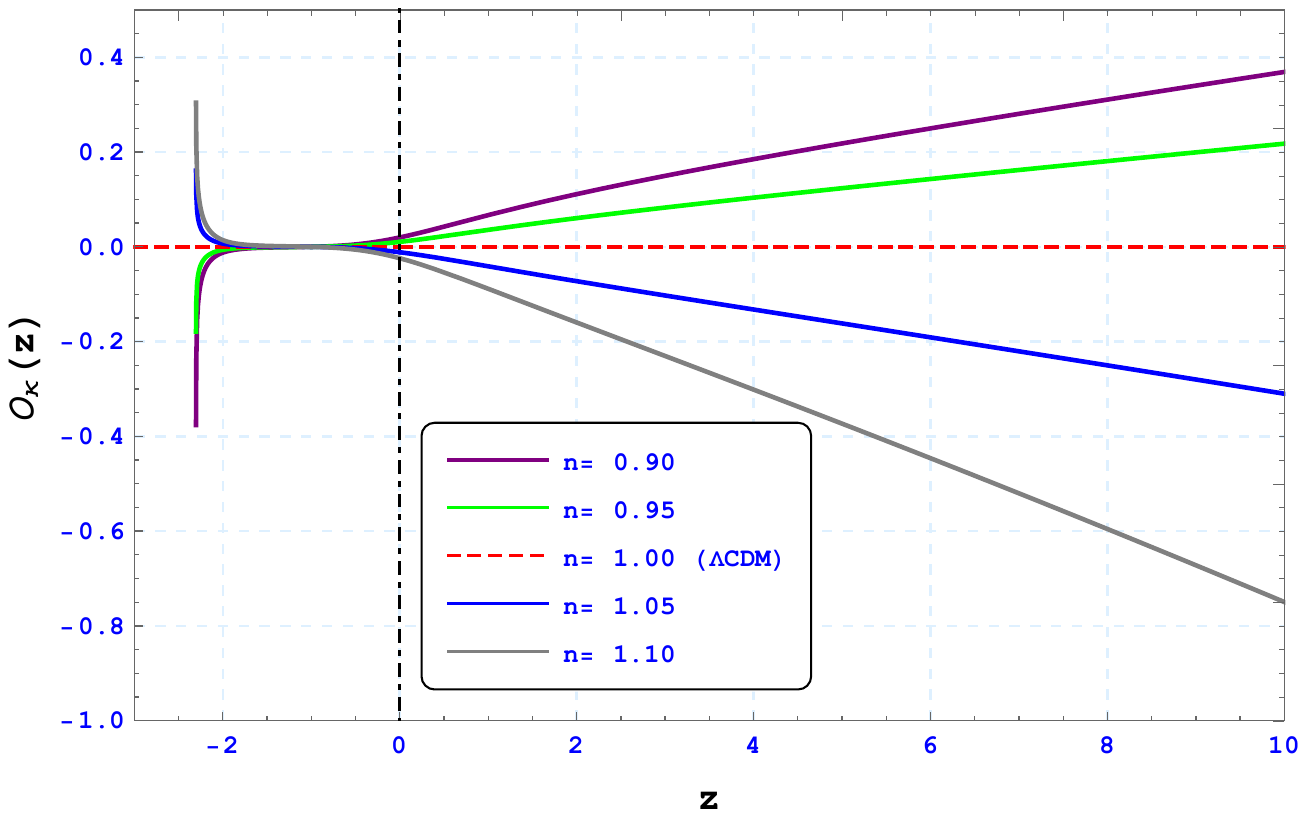}
    \caption{\small(Color online) Plot for $\mathcal{O}_\kappa(z)$ parameter versus redshift $z$ with different model parameter $n$}
    \label{Ok(z).pdf}
\end{figure}
%%%%%%%%%%%%%%%%%%%%%%%%%%%%%%%%%%%%%%%%%%%%%%%%%%%%%%%%%%%%%%%%%%%%%%%%%%%%%%%%%%%%%%%%%%%%%%%%%%%%%%%%%%%%%%%%%%%%%%%%%%%%%%%%%%%%%%%%%
\begin{figure}[htbp!]
    \centering
    \includegraphics[width=10cm,height=8cm]{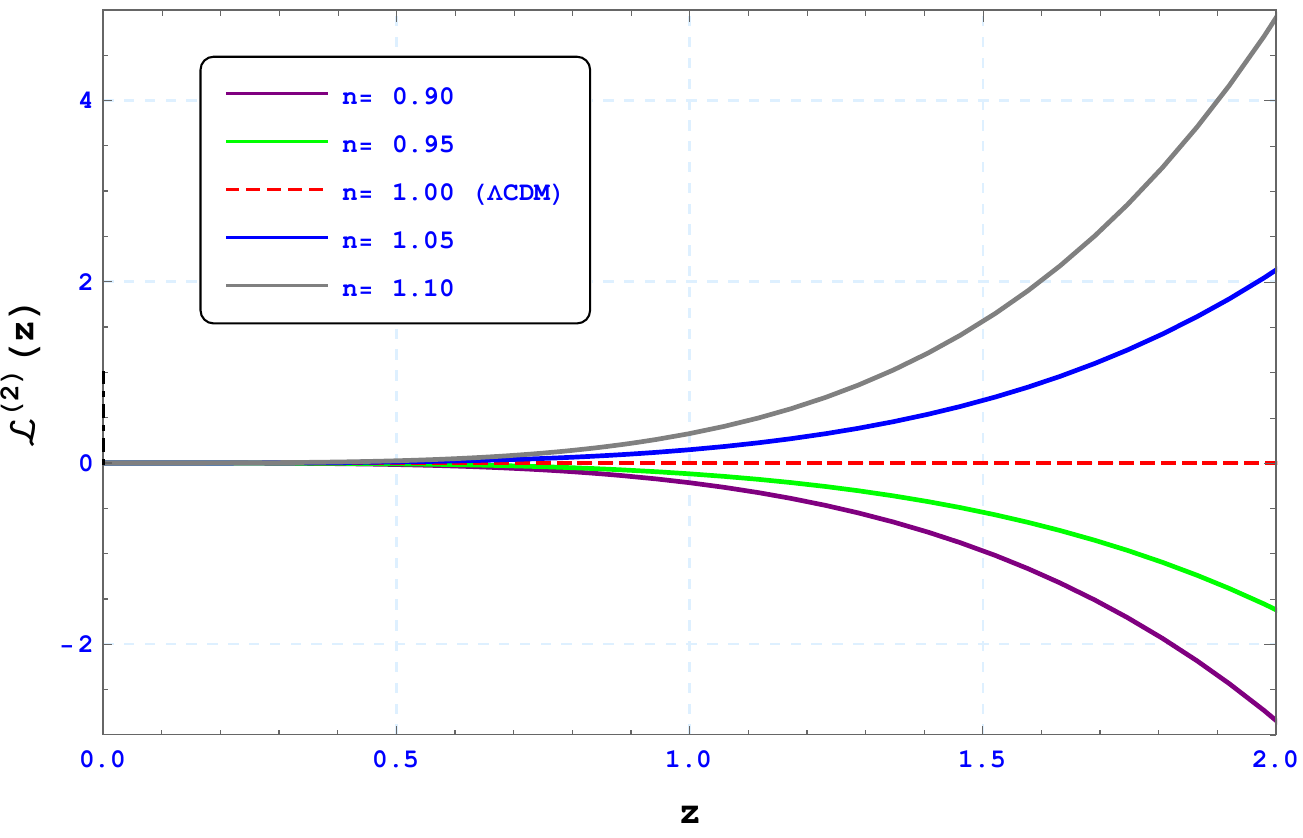}
    \caption{\small(Color online) Plot for $\mathcal{L}^{(2)}(z)$ parameter versus redshift $z$ with different model parameter $n$}
    \label{L2(z).pdf}
\end{figure}
%%%%%%%%%%%%%%%%%%%%%%%%%%%%%%%%%%%%%%%%%%%%%%%%%%%%%%%%%%%%%%%%%%%%%%%%%%%%%%%%%%%%%%%%%%%%%%%%%%%%%%%%%%%%%%%%%%%%%%%%%%%%%%%%%%%%%%%%%
\par Besides cosmography, a novel and well-known geometric diagnostic approach is provided in Refs. \cite{sahni2008two,seikel2012using} to distinguise different cosmological models vis-à-vis flat and non-flat $\Lambda$CDM frameworks. In the GMHE-inspired modified cosmological scenario, one can find the first diagnostic parameter $\mathcal{O}^{(1)}_m(z)$ as
\begin{align}\label{Om(z)_param}
       \mathcal{O}^{(1)}_m(z) &\overset{\text{def}}{=} \frac{E^2(z)-1}{(1+z)^3-1} = \frac{[\Omega_{m,0}(1+z)^{3}+\Omega_{\Lambda,0}]^{\frac{2}{3-n}}-1}{(1+z)^3-1} \,.  
\end{align} $\mathcal{O}^{(1)}_m(z)=\Omega_{m,0}$ implies for the spatially flat $\Lambda$CDM (the concordance model), or in otherwords DE$=\Lambda$. According to Eq. (\ref{Om(z)_param}), we find that $\mathcal{O}^{(1)}_m(z)\big\rfloor_{n = 1} = \Omega_{m,0}$ holds true for any redshift in our model, thereby confirming the fact that $n=1$ returns back to the flat $\Lambda$CDM paradigm. The redshift dependence of $\mathcal{O}^{(1)}_m(z)$ is shown in Fig. \ref{O1m(z).pdf} for various values of model parameter $n$. We observe $\mathcal{O}^{(1)}_m(z)$ is not constant for $n\ne 1$ cases. This signals an alternative DE or modified gravity model, ruling out the standard concordance model for $n\ne 1$ cases. The modification in cosmology influences the constant nature of DE we considered. Whenever $\mathcal{O}^{(1)}_m(z) < \Omega_{m,0}$ (for $n<1$), this model exhibits phantom-like characteristics and $\mathcal{O}^{(1)}_m(z) > \Omega_{m,0}$ (for $n>1$) indicates quintessence-like characteristics. A more effective approach to measure the departure from zero rather than a constant value is by constructing a parameter $\mathcal{L}^{(1)}(z)$ from the first redshift derivative $\mathcal{O}^{(1)\prime}_m(z)$ \cite{zunckel2008consistency,seikel2012using}. In the GMHE-inspired modified cosmological scenario, one yields
\begin{align}
       \mathcal{L}^{(1)}(z) &\overset{\text{def}} {=} 3(1+z)^{2}(1-E(z)^{2}) + 2z(3+3z+z^{2}) E(z) E'(z)\nonumber \\[6pt]
       &= 3(1+z)^2 \left[1+ \left\{ \Omega_{\Lambda 0} - \Omega_{m0} (z+1)^3 \right\}^{\frac{2}{3-n}} +\frac{2z(z^2+3z+3) \Omega_{m0}\left\{\Omega_{m0} (z+1)^3+\Omega_{\Lambda 0} \right\}^{-\frac{1-n}{3-n}}}{(3-n)} \right]\,.
\end{align}
Accordingly, the null test states that $\mathcal{L}^{(1)}(z) \ne 0$ falsifies the concordance model. In Fig. \ref{L1(z).pdf}, we plot $\mathcal{L}^{(1)}(z)$ as a function of redshift for different $n$. It clearly shows non-constant deviation from zero for $n\ne 1$ at all redshifts, supporting the claim of $\mathcal{O}^{(1)}_m$ diagnostics. Therefore, the concordance model is ruled out for $n\ne 1$ from $\mathcal{O}^{(1)}_m$ as well as $\mathcal{L}^{(1)}$ diagnosis. However, it is still possible that a non-flat $\Lambda$CDM model represents the Universe \cite{seikel2012using}. In order to discriminate from the non-flat $\Lambda$CDM model, one can further calculate two diagnostics parameters $\mathcal{O}_{m}^{(2)}(z)$ and $\mathcal{O}_{\kappa}(z)$ in the GMHE-inspired modified cosmological scenario as
\begin{align}
       \mathcal{O}_{m}^{(2)}(z) &\overset{\text{def}}{=} \frac{2 \left[ (1+z) (1-E(z)^2) + z(2+z) E(z) E'(z) \right]}{z^2 (1+z) (3+z)} \nonumber \\[6pt]
       &= \frac{2}{z^2 (3+z)} \left[ 1 - \left\{\Omega_{m0} (1+z)^3 +\Omega_{\Lambda 0}\right\}^{\frac{2}{3-n}} + \frac{3z(1+z)(2+z) \Omega_{m0}\left\{\Omega_{m0} (1+z)^3+\Omega_{\Lambda 0}\right\}^{-\frac{1-n}{3-n}}}{(3-n)} \right]\,,
\end{align}
and
\begin{align}
        \mathcal{O}_{\kappa}(z) &\overset{\text{def}}{=} \frac{3(1+z)^2 [E(z)^2-1] - 2z(3+3z+z^2) E(z) E'(z)}{z^2 (1+z) (3+z)} \nonumber \\[6pt]
        &= \frac{3(1+z)}{z^2(3+z)} \left[-1+ \left\{ \Omega_{\Lambda 0} + \Omega_{m0} (1+z)^3 \right\}^{\frac{2}{3-n}} -\frac{2z(z^2+3z+3) \Omega_{m0}\left\{\Omega_{m0} (1+z)^3+\Omega_{\Lambda 0} \right\}^{-\frac{1-n}{3-n}}}{(3-n)} \right]\,,
\end{align}
respectively. $\mathcal{O}_{m}^{(2)}(z)=\Omega_{m0}$ and $\mathcal{O}_{\kappa}(z)=\Omega_{\kappa}$\footnote{In principle, this is valid for non-flat $\Lambda$CDM model. However, as we work with spatially flat Universe ($\kappa = 0$), $\Omega_{\kappa}=0$ always.} imply non-flat $\Lambda$CDM profile. If these parameters are not consistent with the mentioned standard $\Lambda$CDM values, we can rule out the non-flat $\Lambda$CDM model in principle, and establish that DE$\ne\Lambda$ or there is a modification in gravity. We display $\mathcal{O}_{m}^{(2)}(z)$ and $\mathcal{O}_{\kappa}(z)$ as a function of redhift for different $n$ in Fig. \ref{O2m(z).pdf} and Fig. \ref{Ok(z).pdf}, respectively. We observe non-constant deviations from their $\Lambda$CDM values for $n\ne 1$ cases, suggesting a modified gravity model. Similar to $\mathcal{L}^{(1)}(z)$, a further more effective way to measure the digression from zero rather than a constant value is by defining a parameter $\mathcal{L}^{(2)}(z)$ from the first redshift derivative $\mathcal{O}^{(2)\prime}_m(z)$ or $\mathcal{O^\prime}_{\kappa}(z)$ \cite{zunckel2008consistency,seikel2012using}. In the GMHE-inspired modified cosmological scenario, one finds 
\begin{align}
       \mathcal{L}^{(2)}(z) &\overset{\text{def}}{=} 3(1+z)^{2}(E(z)^{2}-1) - 2z(3+6z+2z^{2}) E(z) E'(z) + z^{2}(3+z)(1+z) \left[ E'(z)^{2} + E(z) E''(z) \right] \nonumber \\[6pt]
       &= 3(1+z)^2 \Bigg[-1+ \left\{ \Omega_{m0} (1+z)^3 +\Omega_{\Lambda 0} \right\}^{\frac{2}{3-n}} + \frac{z\Omega_{m0} \left\{ \Omega_{m0} (1+z)^3 +\Omega_{\Lambda 0} \right\}^{-\frac{2(2-n)}{3-n}}}{(3-n)^2} \nonumber \\
       & \quad \times \Big\{\Omega_{m0}(1+z)^3 \left\{ n(5z^2+15z+6) - 9(1+z)(2+z) \right\} -2(3-n)\Omega_{\Lambda 0}(z^2+3z+3) \Big\} \Bigg]\,.
\end{align}
The null test states that $\mathcal{L}^{(2)}(z) \ne 0$, falsifies the non-flat $\Lambda$CDM model. In Fig. \ref{L2(z).pdf}, we showcase redshift evolution of $\mathcal{L}^{(2)}(z)$ for different $n$. It clearly shows non-constant deviation from zero for $n\ne 1$ at all redshifts, supporting the claim of $\mathcal{O}^{(2)}_m$ as well as $\mathcal{O}_{\kappa}$ diagnostics. Therefore, the non-flat $\Lambda$CDM model is ruled out for $n\ne 1$ from $\mathcal{O}^{(2)}_m$, $\mathcal{O}_{\kappa}$ as well as $\mathcal{L}^{(2)}$ diagnosis. The GMHE-inspired modified cosmology ($n\ne 1$) successfully passes all the litmus tests by falsifying both the flat and non-flat $\Lambda$CDM models.
%%%%%%%%%%%%%%%%%%%%%%%%%%%%%%%%%%%%%%%%%%%%%%%%%%%%%%%%%%%%%%%%%%%%%%%%%%%%%%%%%%%%%%%%%%%%%%%%%%%%%%%%%%%%%%%%%%%%%%%%%%%%%%%%%%%%%%%%%%%%%%%%%%%%%%%%%%%%%%%%%%%%%%%%%%%%
\section{Growth of matter spherical overdensities in GMHE-inspired modified cosmology}\label{Growth of matter spherical overdensities in GMHE-inspired modified cosmology}
\par We assume our background Universe is filled with non-relativistic pressureless dust matter (visible matter + CDM), i.e, $p=p_m=0$, and DE (cosmological constant). The continuity Eq. (\ref{continuity_equation}) for the matter density part reads
\begin{equation}\label{continuity_eqn_background}
    \dot{\rho}_m + 3H\rho_m = 0\,.
\end{equation}
To study the growth of spherical overdensities of matter (perturbation in DE is not considered in our study), a spherically symmetric perturbed cloudy region of radius $a_c$, filled with a dusty homogeneous matter density $\rho^c_m$, is considered. In top-hat SC formalism, this spherical region is described by a uniform density and a top-hat profile, so that at any instant $t$, we write $\rho^c_m(t) = \rho_m(t) + \delta\rho_m(t)$.  If $\delta\rho_m(t) > 0$, this spherical region will ultimately collapse due to its own gravitational instability. If $\delta\rho_m(t) < 0$, it will inflate faster than the average Hubble growth rate, thus creating an underdense region (void) \cite{farsi2022structure,farsi2023evolution,ziaie2020structure}. The consideration of a top-hat profile makes the SC model more convenient, as the uniformity of the perturbation is preserved throughout the collapse process, resulting in an exclusively time-dependent evolution, rather than a space-dependent one. Consequently, we can avoid gradients inside the perturbed regime. The top-hat SC model effectively illustrates the evolution of a uniform mini-Universe within a larger, uniform Universe \cite{fernandes2012spherical}. In this context, it is worth mentioning that, during the matter-dominated era of the Universe, the growth of overdense regions slows down compared to the rest of the Universe. This suggests that if their density becomes sufficiently high, ultimately they collapse into clusters and other gravitationally bound systems \cite{Ryden:1970vsj}. Similar to Eq. (\ref{continuity_eqn_background}), the continuity equation for the matter inside the spherically perturbed cloud with radius $a_c$ and local expansion rate $H_c=\frac{\dot a_c}{a_c}$, takes the form
\begin{equation}\label{continuity_eqn_peturbation}
    \dot{\rho}^c_m + 3H_c\rho^c_m = 0\,.    
    \end{equation}
Now, to investigate the growth of perturbation, we start by defining matter density contrast (MDC), a dimensionless quantity
    \begin{equation}\label{density_contrast}
    \delta_m = \frac{\rho^c_m - \rho_m}{\rho_m}
    = \frac{\delta \rho_m}{\rho_m} \, ,
    \end{equation}
measuring the difference in matter densities of the local and background fluids. The first and second-order derivatives of MDC with respect to cosmic time read
    \begin{equation}\label{first derivative_density contrast}
    \dot{\delta}_m = 3(1+\delta_m)(H - H_c) \,,
    \end{equation}
    \begin{equation}\label{second derivative_density contrast}
    \ddot{\delta}_m
     = 3(\dot{H} - \dot{H_c})(1+\delta_m)
    + \frac{\dot{\delta}_m^{\,2}}{1+\delta_m} \,.
\end{equation}
In deriving the above Eqs. (\ref{first derivative_density contrast}) and (\ref{second derivative_density contrast}), we have used Eqs. (\ref{continuity_eqn_background}) and (\ref{continuity_eqn_peturbation}). Combining Eqs. (\ref{2nd_Friedmann_eqn}), (\ref{mod_Friedmann_eqn}), and using $\dot{H}=\frac{\ddot{a}}{a} - H^2$, we obtain for the background
\begin{equation}\label{background_evolution}
    \frac{\ddot{a}}{a} = \frac{-n\, \Gamma_{\gamma,n}^\frac{2}{3-n}}{3-n}(\rho_m + \rho_\Lambda)^\frac{2}{3-n} + \frac{3\, \Gamma_{\gamma,n}^\frac{2}{3-n}}{3-n}(\rho_m + \rho_\Lambda)^{-\frac{1-n}{3-n}}\rho_\Lambda \, .
\end{equation}
In accordance with SC formalism, a homogeneous spherically perturbed region with radius $a_c$ can itself be described by the same equations that dictate the evolution of the background Universe characterized by scale factor $a$ \cite{peebles2020principles}. Thus, for the spherically perturbed cloud with a radius $a_c$, it follows similar form to Eq. (\ref{background_evolution}), specifically
\begin{equation}\label{perturbed_evolution}
    \frac{\ddot{a}_c}{a_c} = \frac{-n\, \Gamma_{\gamma,n}^\frac{2}{3-n}}{3-n}(\rho^c_m + \rho_\Lambda)^\frac{2}{3-n} + \frac{3\, \Gamma_{\gamma,n}^\frac{2}{3-n}}{3-n}(\rho^c_m + \rho_\Lambda)^{-\frac{1-n}{3-n}}\rho_\Lambda \, .
\end{equation}
In general, one may anticipate $\gamma$, $n$, and $\rho_\Lambda$ (or $\Lambda$) are different for background and perturbed regions. However, we assume they are the same, i.e. $\gamma_c = \gamma$, $n_c = n$, and $\rho^c_\Lambda = \rho_\Lambda$ (or $\Lambda_c = \Lambda$), to avoid complexity in the system. The term $(\dot{H} - \dot{H_c})$ in Eq. (\ref{second derivative_density contrast}) can be evaluated using Eqs. (\ref{density_contrast}), (\ref{background_evolution}), and (\ref{perturbed_evolution}) as
\begin{align}\label{Hdot-hdot}
    \dot{H} - \dot{H_c} &= -H^2 + H_c^2 + \frac{2n\, \Gamma_{\gamma,n}^\frac{2}{3-n}}{(3-n)^2}\frac{\rho_m}{(\rho_m + \rho_\Lambda)^{\frac{1-n}{3-n}}}\delta_m  
    + \frac{3(1-n)\, \Gamma_{\gamma,n}^\frac{2}{3-n}}{(3-n)^2}\frac{\rho_m\,\rho_\Lambda}{(\rho_m + \rho_\Lambda)^{\frac{2(2-n)}{3-n}}}\delta_m \, .
\end{align}
Since, we are working within the linear regime where $\delta_m <1$, we neglect $\mathcal{O}(\delta^2_m)$ and $\mathcal{O}(\dot{\delta}^2_m)$ in evaluating Eq. (\ref{Hdot-hdot}). With the same argument, the last term in Eq. (\ref{second derivative_density contrast}) can also be neglected. By merging Eq. (\ref{Hdot-hdot}) with Eq. (\ref{second derivative_density contrast}), and making use of Eq. (\ref{first derivative_density contrast}), the linear differential equation representing the temporal evolution of MDC yields
\begin{align}\label{temoral evolution eqn}
\ddot{\delta}_m
+ 2H \dot{\delta}_m
- \frac{6n\, \Gamma_{\gamma,n}^{\frac{2}{3-n}}}{(3-n)^2}
\frac{\rho_m}{(\rho_m + \rho_\Lambda)^{\frac{1-n}{3-n}}}\,\delta_m
- \frac{9(1-n)\, \Gamma_{\gamma,n}^{\frac{2}{3-n}}}{(3-n)^2}
\frac{\rho_m\,\rho_\Lambda}
{(\rho_m + \rho_\Lambda)^{\frac{2(2-n)}{3-n}}}\,\delta_m
= 0 \, .
\end{align}
To study the evolution of MDC concerning the redshift parameter, we first substitute the time derivatives with the derivatives with respect to the scale factor. It is easy to show that
\begin{subequations}\label{time to scale factor derivatives}
    \begin{align}
       \dot{\delta}_m &= aH\delta^\prime_m \,,\\
       \ddot{\delta}_m &= aH^2\delta^{\prime\prime}_m + a\left(\frac{\ddot{a}}{a}\right)\delta^\prime_m\,, 
    \end{align}
\end{subequations}
where primes denote the order of derivatives with respect to the scale factor $a$. Therefore, employing Eqs. (\ref{Hubble}), (\ref{background_evolution}) and (\ref{time to scale factor derivatives}) in Eq. (\ref{temoral evolution eqn}), we obtain
\begin{align}\label{scale factor evolution equation_one}
\delta_m^{\prime\prime}
+ \frac{3}{(3-n)a}
\left\{
(2-n)
+ \frac{\Gamma_{\gamma,n}}{H^{3-n}}\,\rho_\Lambda
\right\}
\delta_m^{\prime}
- \frac{6n}{(3-n)^2}
\frac{\Gamma_{\gamma,n}}{a^2 H^{3-n}}\,
\rho_m\,\delta_m
- \frac{9(1-n)}{(3-n)^2}
\frac{\Gamma_{\gamma,n}^2}{a^2 H^{2(3-n)}}\,
\rho_m\,\rho_\Lambda\,\delta_m
= 0 \, .
\end{align}
The Eq. (\ref{scale factor evolution equation_one}) provided above can be further simplified by incorporating the density parameters outlined in Eq. (\ref{density_parameters}), which yields
\begin{align}\label{scale factor evolution equation_two}
\delta_m^{\prime\prime}
+ \frac{3}{(3-n)a}
\left\{
(2-n) + (1-\Omega_m)
\right\}
\delta_m^{\prime}
- \frac{6n}{(3-n)^2 a^2}\,
\Omega_m\,\delta_m
- \frac{9(1-n)}{(3-n)^2 a^2}\,
\Omega_m\,(1-\Omega_m)\,\delta_m
= 0 \, .
\end{align}
It is interesting to note that the multiplicative parameter $\gamma$ is absent in the MDC evolution equation, whereas the entropic exponent parameter $n$ significantly influences it. For a pure matter-dominated GMHE-inspired Universe ($\Omega_m \approx 1$ and $\Omega_\Lambda \approx 0$), the Eq. (\ref{scale factor evolution equation_two}) reduces to
\begin{equation}\label{matter_domination_GMHE}
    \delta^{\prime\prime}_m + \frac{3(2-n)}{(3-n)a}\delta^{\prime}_m - \frac{6n}{(3-n)^2a^2}\delta_m = 0\, ,
\end{equation}
with a solution expressed as a function of redshift
\begin{equation}\label{sol_matter_dom_GMHE}
    \delta_m(z) = \mathcal{A}_n (1+z)^{-\frac{2n}{3-n}} + \mathcal{B}_n (1+z)^{\frac{3}{3-n}}\, ,
\end{equation}
which has been explored recently in Ref. \cite{luciano2025modified_Cos}. Here $\mathcal{A}_n$, and $\mathcal{B}_n$ are integration constants. Once again, one can observe that, in the limit $n=1$, Eq. (\ref{matter_domination_GMHE}) reduces to
\begin{equation}\label{matter_domination_SC}
    \delta^{\prime\prime}_m + \frac{3}{2a}\delta^{\prime}_m - \frac{3}{2a^2}\delta_m = 0\, ,
\end{equation}
that admits a widely known solution in terms of redshift
\begin{equation}\label{sol_matter_dom_GR}
    \delta_m(z) = \mathcal{A}_1 (1+z)^{-1} + \mathcal{B}_1 (1+z)^{\frac{3}{2}}\, ,
\end{equation}
coinciding with the result obtained in pure matter-domination in the standard cosmology (the GR limit) \cite{padmanabhan1993structure, weinberg2013gravitation,abramo2007structure}. This suggests that the presence of DE under the influence of GMRE-inspired cosmology shall also play a significant role in understanding the evolution of perturbations, in addition to the dynamics of this modified cosmology. As we wish to analyze MDC in terms of redshift, we exploit a set of transformations
\begin{subequations}\label{scale factor to redshift derivatives}
    \begin{align}
       \delta^\prime_m &= -(1+z)^2\, \frac{d\delta_m}{dz} \,,\\
       \delta^{\prime\prime}_m &= (1+z)^4\,\frac{d^2\delta_m}{dz^2}+2(1+z)^3 \,\frac{d\delta_m}{dz}\,, 
    \end{align}
\end{subequations}
into Eq. (\ref{scale factor evolution equation_two}), which gives
\begin{align}\label{evolution eqn}
(1+z)^2\,\frac{d^2 \delta_m}{dz^2}
+ (1+z)
\left\{
\frac{n - 3(1-\Omega_m)}{(3-n)}
\right\}
\frac{d \delta_m}{dz}
- \frac{3}{(3-n)^2}
\left\{
2n\,\Omega_m
+ 3(1-n)(1-\Omega_m)\,\Omega_m
\right\}
\delta_m
= 0 \, .
\end{align}
Considering the dynamical behavior of the cosmological matter density parameter provided in Eq. (\ref{matter_density_parameter}), Eq. (\ref{evolution eqn}) admits the following analytical solution
\begin{align}\label{sol_TOTAL_GMHE}
\delta_m(z)
&= \mathcal{C}_{n}\,
\left[
\Omega_{m,0}(1+z)^{3}
+ \Omega_{\Lambda,0}
\right]^{\frac{1}{3-n}}
+ \frac{\mathcal{D}_{n}}{2}\,(1+z)^{2}
\left(
1 + \frac{\Omega_{m,0}(1+z)^{3}}{\Omega_{\Lambda,0}}
\right)^{\frac{3}{3-n}}
\left[
\Omega_{m,0}(1+z)^{3}
+ \Omega_{\Lambda,0}
\right]^{-\frac{2}{3-n}}
\nonumber \\[4pt]
&\quad\times
{}_{2}\mathfrak{F}_{1}\!\left(
\frac{2}{3},\,
\frac{3}{(3-n)};\,
\frac{5}{3};\,
-\frac{\Omega_{m,0}(1+z)^{3}}{\Omega_{\Lambda,0}}
\right)\, .
\end{align}
where $\mathcal{C}_{n}$ and $\mathcal{D}_{n}$ are integration constants, and a function of form $ {}_{2}\mathfrak{F}_{1}(a,b;c;z)$ is represented as a hypergeometric function with parameters $a,b$, and $c$, and a variable $z$. For the $\Lambda$CDM model with $n=1$, the above solution reduces to
\begin{align}\label{delta_solution_SC}
\delta_m(z)
&= \mathcal{C}_{1}\,
\sqrt{
\Omega_{m,0}(1+z)^3 + \Omega_{\Lambda,0}
}
+ \frac{\mathcal{D}_{1}\,(1+z)^2}{2\,\Omega_{\Lambda,0}}
\sqrt{1+
\frac{\Omega_{m,0}(1+z)^3}{\Omega_{\Lambda,0}}
} \times
{}_{2}\mathfrak{F}_{1}\!\left(
\frac{2}{3},\,
\frac{3}{2};\,
\frac{5}{3};\,
-\frac{\Omega_{m,0}(1+z)^3}{\Omega_{\Lambda,0}}
\right)\, .
\end{align}
Since our analysis is restricted to the linear regime and focused on pre-collapsed evolution of perturbations, to obtain the integration constants of the solution given in Eq. (\ref{sol_TOTAL_GMHE}), we adopt adiabatic initial conditions 
\begin{equation}\label{initial_conditions}
    \delta_m(z)\big\rfloor_{z = z_i}=\delta^i_m \quad\text{and} \quad \frac{d\delta_m(z)}{dz}\Bigg\rfloor_{z = z_i} = -\left(\frac{2n}{3-n}\right)\frac{\delta^i_m}{1+z_i}\, .
\end{equation}
%%%%%%%%%%%%%%%%%%%%%%%%%%%%%%%%%%%%%%%%%%%%%%%%%%%%%%%%%%%
\begin{figure}[htbp!]
    \centering
    \includegraphics[width=10cm,height=8cm]{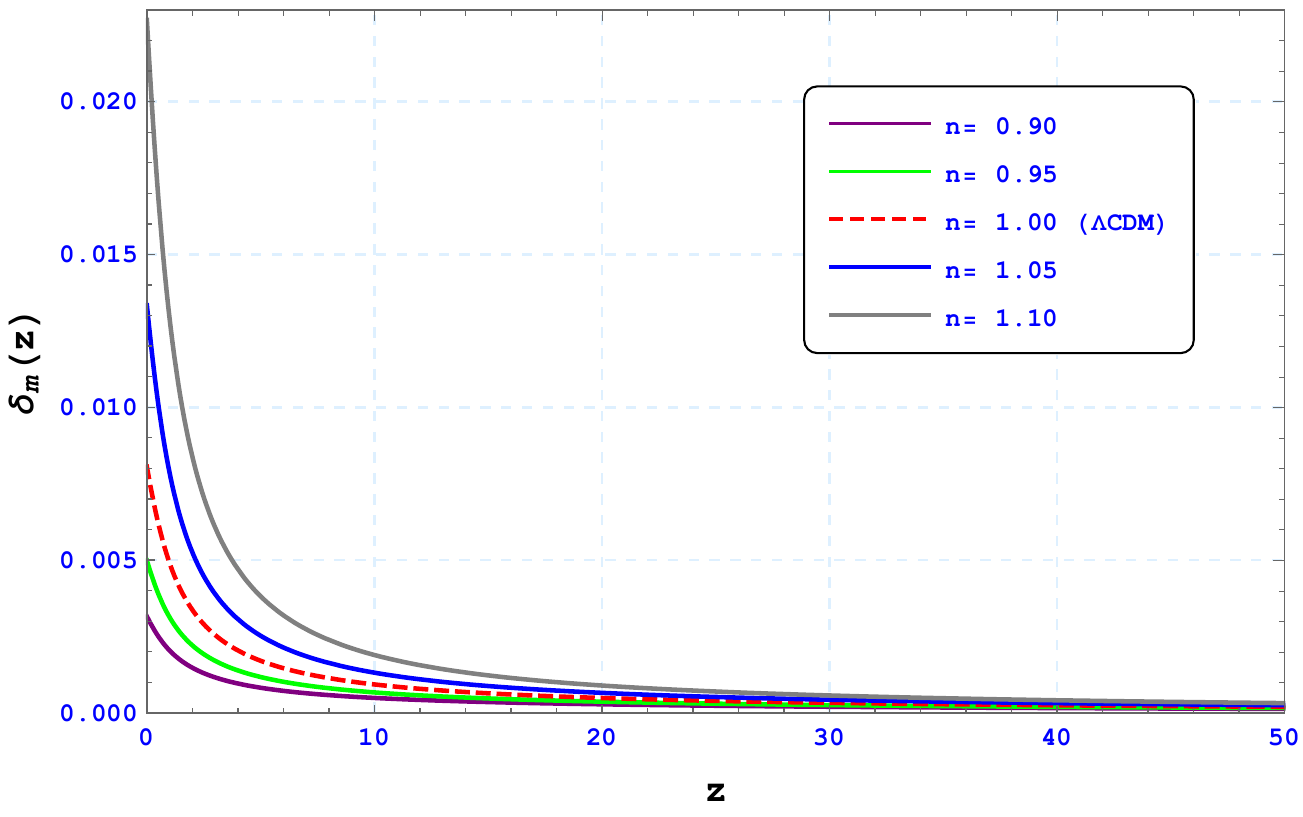}
    \caption{\small(Color online) Plot for matter density contrast $\delta_m(z)$ parameter versus redshift $z$ with different model parameter $n$}
    \label{delta_m(z).pdf}
\end{figure}
%%%%%%%%%%%%%%%%%%%%%%%%%%%%%%%%%%%%%%%%%%%%%%%%%%%%%%%%%%%$
In particular, we select $z_i = 10^3$, corresponding to an epoch just after matter-radiation equality and well before recombination, and set $\delta^i_m = 10^ {-5}$ \cite{luciano2025modified_Cos}, which aligns with the expected scale of primordial fluctuations as predicted by inflation theory, and is corroborated by the CMBR anisotropy measurements from the Planck mission \cite{aghanim2020planck,aghanim2021erratum}. The initial condition on the first-order derivative of the MDC comes from the fact that DE is negligible initially at the matter-domination, shortly after matter-radiation equality. To obtain the expression for the same, we take only the growing part of the solution, which is important for the structure formation,  given in Eq. (\ref{sol_matter_dom_GMHE}) by considering $\mathcal{B}_n = 0$. In this context, we state that we consider only minor deviations from the standard $\Lambda$CDM model, mathematically represented as $|n-1|<<1$. In Fig. \ref{delta_m(z).pdf}, we have plotted MDC $\delta_m$ as a function of redshift $z$ for various values of the entropic exponent parameter $n$. It is evident that the influence of GMHE-inspired modified cosmology leaves a noticeable impression on the nature of the MDC. We observe that the MDC starts to grow from the beginning and grows rapidly as the Universe expands, showing a difference from the $\Lambda$CDM profile ($n=1$). We also observe that the growth of perturbations increases with increasing $n$, and particularly in the lower redshifts, the impact of parameter $n$ is clearly distinguishable. For $n<1$, the MDC grows at a slower rate compared to standard cosmology, while for $n>1$ it grows at a faster rate.
\par
\begin{figure}[htbp!]
    \centering
    \includegraphics[width=10cm,height=8cm]{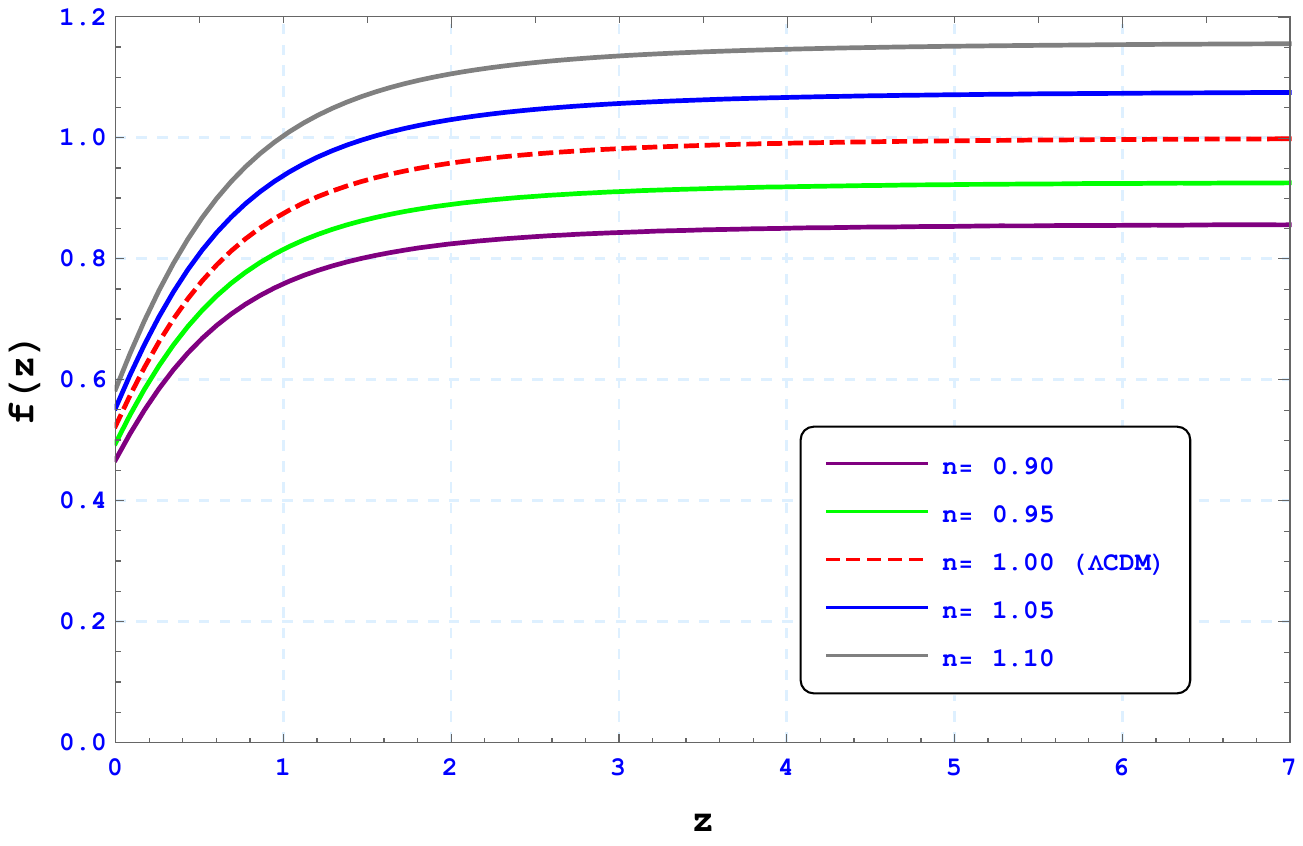}
    \caption{\small(Color online) Plot for logarithmic growth rate $f(z)$ parameter versus redshift $z$ with different model parameter $n$}
    \label{f(z).pdf}
\end{figure}

 In this context, we can investigate the rate of growth of matter perturbations through the logarithmic growth function (LGF) \cite{peebles2020principles}
\begin{equation}\label{logarithmic growth function}
    f(z) \overset{\text{def}}{=} -\frac{d\, ln \delta_m(z)}{d\, ln(1+z)} = -(1+z)\frac{1}{\delta_m(z)}\frac{d\,\delta_m(z)}{dz}\, .
\end{equation}
We have plotted the LGF in terms of redshift for different $n$ values, and showed the $\Lambda$CDM profile in contrast in Fig. \ref{f(z).pdf}. We observe that the value of the LGF approaches a saturation value (unity in standard $\Lambda$CDM cosmology) at high redshifts, whereas at lower redshifts it starts to decrease. The overall and current value of $f(z)$ firmly depend on the model parameter $n$, and increase with increasing $n$. The decrease of LGF at the lower reshifts is due to the DE domination, which slows down the structure formation by suppressing the growth. 
\par
\begin{figure}[htbp!]
    \centering
    \includegraphics[width=10cm,height=8cm]{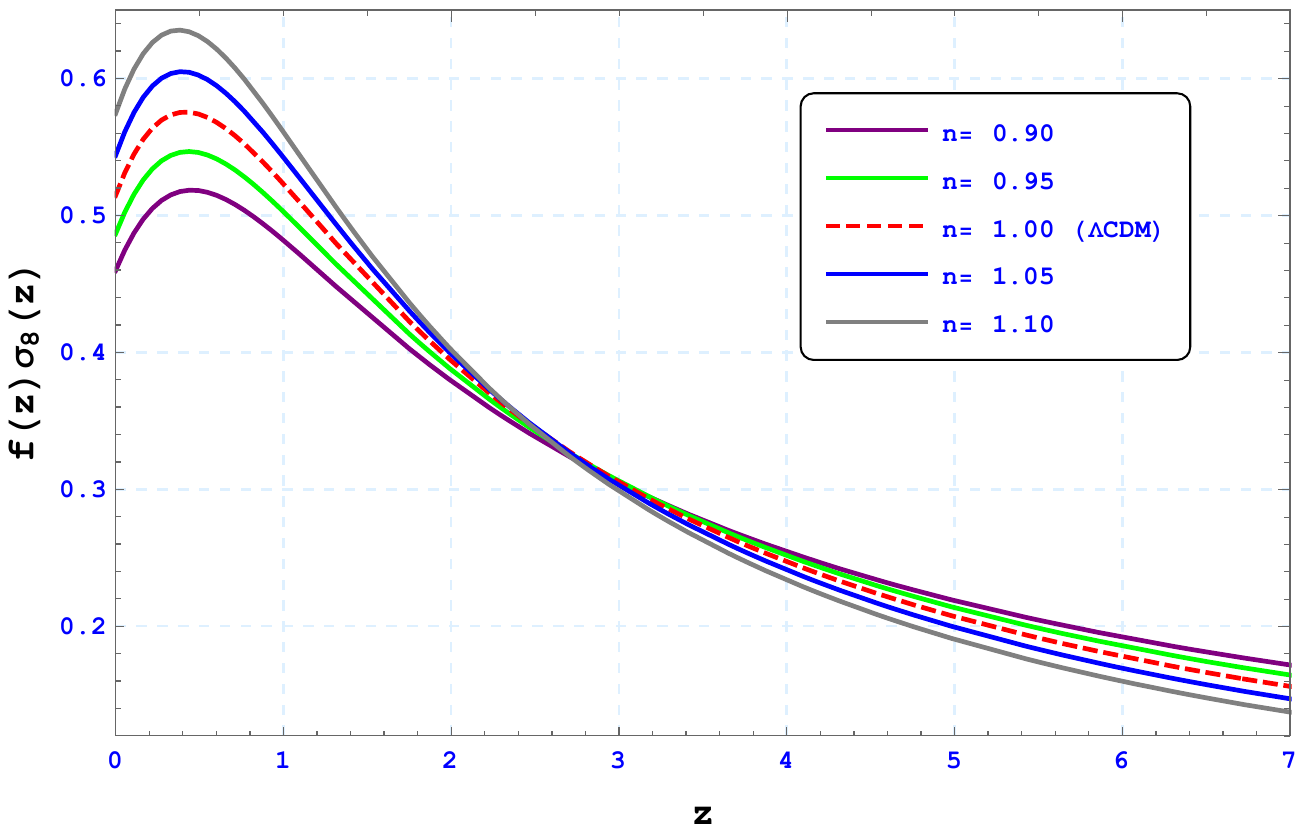}
    \caption{\small(Color online) Plot for $f(z) \sigma_8(z)$ parameter versus redshift $z$ with different model parameter $n$}
    \label{f(z).sigma_8(z).pdf}
\end{figure}

 There is also an assessment of the growth rate of matter fluctuations by inspecting the redshift-space distortion in the clustering arrangement of galaxies. This distortion stems from the peculiar velocities associated with the inward collapse of large-scale structures, which are directly associated with the growth rate of the MDC \cite{kaiser1987clustering}. Recent serveys exploring galaxy redshift has helped to establish bounds on the LGF $f(z)$ or $f(z)\sigma_8(z)$ as a function of redshift, where $f(z)$ is from Eq. (\ref{logarithmic growth function}) and $\sigma_8(z)$ represents the root mean squared amplitude of $\delta_m$ measured at a comoving scale $8 \mathrm{h}^{-1} Mpc$ \cite{tsujikawa2013testing,nesseris2017tension}, and can be expressed as \cite{nesseris2008testing}
\begin{equation}
    \sigma_8(z) \overset{\text{def}}{=} \frac{\delta_m(z)}{\delta_m(z=0)}\sigma_8(z=0)
\end{equation}
where we have supposed $\sigma_8(z=0)=0.983 \pm 0.0060$ \cite{nesseris2017tension}. The redshift evolution of $f(z)\sigma_8(z)$ for different values of the $n$ parameter is also visualized in Fig. \ref{f(z).sigma_8(z).pdf}. In the case of small redshifts ($z<1$), this GMHE-inspired modified cosmological model with a larger $n$ features a larger value of the cosmological growth rate. However, at large redshifts, the behavior turns different. Specifically, by incorporating nonlinearity into SMHR within the GMHE-inspired modified cosmological framework, we find that at low redshifts the growth rate function surpasses that of the $\Lambda$CDM model for $n>1$, while it falls below for $n<1$. Moreover, with the inclusion of model parameter $n$, we also observe that $f(z)\sigma_8(z)$ achieves its peak value at lower redshifts for larger $n$ values. In other words, the large-scale structures form later in the GMHE-inspired modified cosmological model than in the standard $\Lambda$CDM counterpart for $n>1$; conversely, they form at an earlier epoch for $n<1$. This insight illustrates the significant impact of the model parameter $n$ on the timing of structure formation in the Universe.
%%%%%%%%%%%%%%%%%%%%%%%%%%%%%%%%%%%%%%%%%%%%%%%%%%%%%%%%%%%
%%%%%%%%%%%%%%%%%%%%%%%%%%%%%%%%%%%%%%%%%%%%%%%%%%%%%%%%%%%
%%%%%%%%%%%%%%%%%%%%%%%%%%%%%%%%%%%%%%%%%%%%%%%%%%%%%%%%%%%%%%%%%%%%%%%%%%%%%%%%%%%%%%%%%%%%%%%%%%%%%%%%%%%%%%%%%%%%%%%%
\section{Halo mass function and cluster number counts in GMHE-inspired modified cosmology}\label{Halo mass function and cluster number counts in GMHE-inspired modified cosmology}
\par Apart from the redshift evolution of the MDC and growth functions, it would be interesting and worthy to survey the number counts of the collapsed objects in the GMHE-inspired modified cosmological scenario in this article. We know that the underlying mechanism of large-scale structure formation in the Universe is the gravitational collapse of matter perturbations. The collapsed objects are called the DM halos. The visible baryonic matter follows the DM distribution due to gravitational interaction. Every galaxy cluster, including the cluster containing our Milky Way, is believed to be embedded within one of these hefty and roughly spherical clouds or halos of DM. The observation of the distribution of galaxy clusters leads to information about the distribution of halo structures of DM in the Universe. In this section, we study the \textit{differential halo mass function} (DHMF) and cluster number counts of DM halos in the GMHE-inspired modified cosmological framework through the Sheth-Mo-Tormen (SMT) \cite{sheth1999large,sheth2001ellipsoidal,sheth2002excursion} formalism, an improved version of the Press-Schechter (PS) \cite{press1974formation} method. 
\par Using the present mean mass density $\rho_{m,0}$, root-mean-square of smoothed density fluctuation $\sigma(M,z)$ in a sphere of mass $M$ and comoving radius $R$, and critical overdensity $\delta_{cr}(z)$ above which structures collapse, we compute the comoving number density of collapsed halos per logarithmic interval (a.k.a DHMF) in the range of virial mass $M$ to $M+dM$ at a given redshift z \cite{herrera2017calculation,gupta2022universality}
\begin{equation}\label{DHMF}
    \frac{dn}{d \ln M}(M,z) =  \frac{\rho_{m,0}}{M}  \left| \frac{d \ln \sigma(M,z)}{d \ln M} \right|\mathcal{F}(M,z)\, ,
\end{equation}
where the \textit{halo multiplicity function} $\mathcal{F}(M,z)$ determines the shape of the DHMF or the mass fraction in the collapsed volume. The DHMF measures the comoving number density of DM halos. The abundance of DM halos is described as a function of virial halo mass at a certain redshift, subjected to a unit volume normalization. The value of the DHMF relies on the cosmological background Universe as well as the matter power spectrum. A primitive theoretical model of this statistical quantity was formulated by Press and Schechter \cite{press1974formation}, based on a basic theory of the hierarchical Gaussian density fluctuation field with a fixed barrier. The most original version of the PS model was modified later by using the excursion set method, also known as the extended PS formalism (see Ref. \cite{bond1991excursion}), which proposed the functional form of $\mathcal{F}(M,z)$ as follows
\begin{equation}
    \mathcal{F}_{PS}(M,z) = \sqrt{\frac{2}{\pi}}\frac{\delta_{cr}(z)}{\sigma(M,z)}   \exp\left( -\frac{\delta_{cr}^2(z)}{2\sigma^2 (M,z)}\right)\, .
\end{equation}
%%%%%%%%%%%%%%%%%%%%%%%%%%%%%%%%%%%%%%%%%%%%%%%%%%%%%%%%%%%
\begin{figure}[htbp!]
    \centering
    \includegraphics[width=10cm,height=8cm]{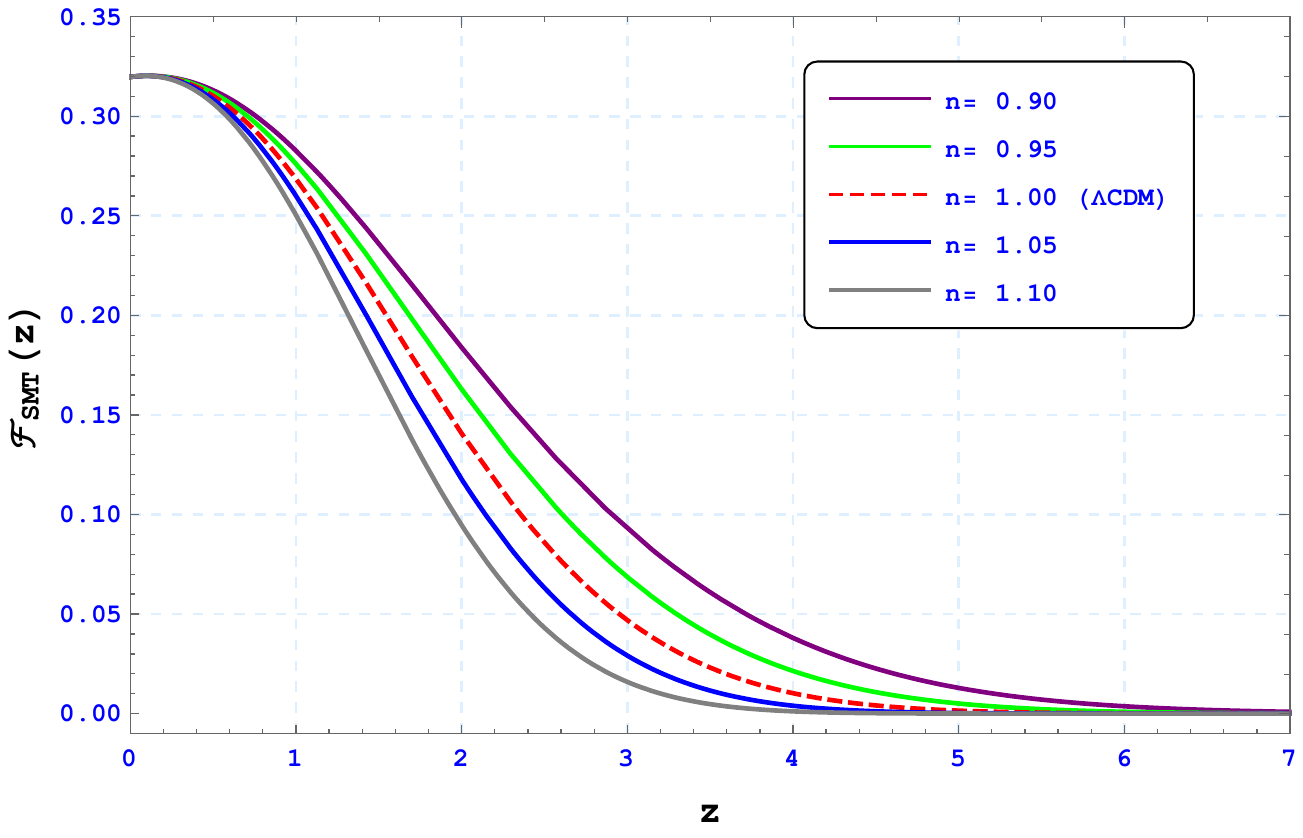}
    \caption{\small(Color online) Plot for Sheth, Mo, and Torman halo multiplicity function $\mathcal{F}_{\mathrm{SMT}}$ versus redshift $z$ for mass $M=10^{13}\mathrm{h}^{-1}M_{\odot}$ with different model parameter $n$}
    \label{f_SMT(z).pdf}
\end{figure}
%%%%%%%%%%%%%%%%%%%%%%%%%%%%%%%%%%%%%%%%%%%%%%%%%%%%%%%%%%%
%%%%%%%%%%%%%%%%%%%%%%%%%%%%%%%%%%%%%%%%%%%%%%%%%%%%%%%%%%%
\begin{figure}[htbp!]
    \centering
    \includegraphics[width=10cm,height=8cm]{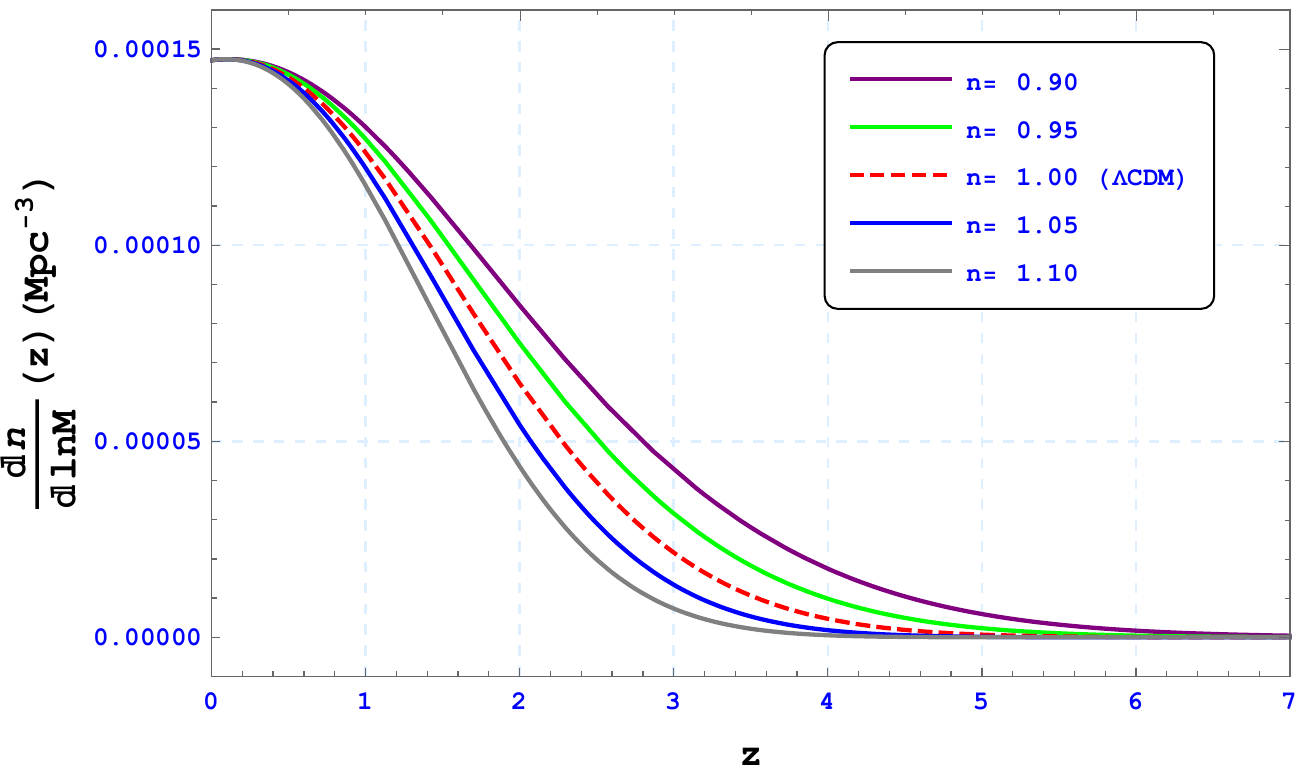}
    \caption{\small(Color online) Plot for Sheth, Mo, and Torman--corrected differential halo mass function $\frac{dn}{d \ln M}(M,z)\big\rfloor_{SMT}$ versus redshift $z$ for mass $M=10^{13}\mathrm{h}^{-1}M_{\odot}$ with different model parameter $n$}
    \label{dnBYdlnM(z).pdf}
\end{figure}
%%%%%%%%%%%%%%%%%%%%%%%%%%%%%%%%%%%%%%%%%%%%%%%%%%%%%%%%%%%
The aforementioned model has proven remarkably successful, as it shows fair unanimity with the results produced from simulation, particularly in the moderate halo mass regime. The advent of precision cosmology, along with the swift development of resolution and size of N-body simulations, enabled a robust prediction of numerical values of DHMF over several orders of magnitude \cite{watson2013halo,tinker2008toward,crocce2010simulating,courtin2011imprints,bond1991excursion,hellwing2016copernicus,pillepich2018first}. These have suggested that original $\mathcal{F}_{PS}(M,z)$ significantly overestimates the number of low-mass halos, while underpredicting the number of massive halo structures in the regime of cluster mass. In this context, several reformed models have been proposed in the literature (see Refs. \cite{tinker2008toward,crocce2010simulating,courtin2011imprints, sheth2001ellipsoidal,lukic2007halo,peacock1990alternatives,reed2013towards,watson2013halo}). Although all of these alternative models have their own benefits/drawbacks \cite{Murray:2013qza}, it has been tested that most of the modified DHMF models achieve comparable accuracy \cite{gupta2022universality}. Therefore, for the sake of simplicity and brevity, we pick one specific model that allows for an ellipsoidal peak shape, along with the possibility of a moving barrier, while relaxing PS suppositions for our analytical predictions
\begin{align}\label{F_SMT}
\mathcal{F}_{\mathrm{SMT}}(M, z) 
&= A \sqrt{\frac{2a}{\pi}}
\left[
1 + \left(
\frac{\sigma^2(M,z)}{a \delta_{cr}^2(z)}
\right)^p
\right]
\frac{\delta_{cr}(z)}{\sigma(M,z)}
\exp\left(
-\frac{a \delta_{cr}^2(z)}
{2 \sigma^2(M,z)}
\right)\, ,
\end{align}
put forward by Sheth, Mo, and Torman in Ref. \cite{sheth2001ellipsoidal}, where $p=0.3$, $a=0.707$, and $A\approx 0.3222$ were obtained from the simulation of DM halo formation. Theoretically, the critical density threshold $\delta_{cr}(z)$ depends on redshift and the parameters of the governing modified cosmology. However, it has been observed that $\delta_{cr}(z)$ exhibits less sensitivity to their changes when analyzed across different modified cosmologies, especially after a few redshifts \cite{vianapedro1996cluster,herrera2017calculation,nunes2006structure,velten2014structure,abramo2007structure,farsi2022structure,farsi2023evolution,gupta2022universality,mukherjee2025spherical}. Additionally, we checked that DHMF shows no appreciable change when slightly varying from $\delta_{cr}=1.686$ \cite{padmanabhan1993structure}, which corresponds to the Einstein-de Sitter (and the $\Lambda$CDM) Universe. Thus, to avoid extra model-dependent complexity, we shall utilize the standard $\Lambda$CDM spherical collapse based $\delta_{cr}=1.686$ for obtaining all the analytical predictions in the modified cosmological framework inspired by GMHE with parameter ($n$) values that fluctuate to some degree around $n=1$, corresponding to the $\Lambda$CDM scenario. The true shape of $\sigma(M,z)$ uses to be approximated around a fixed length $R_8 = 8h^{-1}\, Mpc$, corresponding to mass $M_8=5.95\times10^{14}\,\Omega_{m,0}\mathrm{h}^{-1}M_{\odot}$ \cite{herrera2017calculation} ($M_{\odot}$ is the solar mass) and we use the fitting from Ref. \cite{vianapedro1996cluster}
\begin{equation}
    \sigma(M,z)=\sigma(M_8,0)\frac{\delta_m(z)}{\delta_m(z=0)}\left(\frac{M}{M_8}\right)^{-\frac{\gamma (M)}{3}}\, ,
\end{equation}
where 
\begin{equation}
    \gamma (M) = (0.3\,\mathrm{h}\,\Gamma+0.2)\left[2.92+\frac{1}{3}\log\left(\frac{M}{M_8}\right)\right]\,,
\end{equation}
and the shape parameter is given by \cite{sugiyama1994cosmic,vianapedro1996cluster}
\begin{equation}
    \Gamma = \Omega_{m,0}\,\mathrm{h}\exp\left[-\Omega_{bar,0}\left(1+\frac{1}{\Omega_{m,0}}\right)\right]\,,
\end{equation}
with the current baryon density $\Omega_{bar,0} = 0.02230/\mathrm{h}^2$ \cite{herrera2017calculation}. Using the above expressions, Eq. (\ref{DHMF}) can be rewritten by inducing SMT corrections as
\begin{align}\label{SMT_DHMF}
\frac{dn}{d \ln M}(M,z)\bigg\rfloor_{SMT} &=  \frac{\rho_{m,0}}{M}  \left| \frac{d \ln \sigma(M,z)}{d \ln M} \right|
A \sqrt{\frac{2a}{\pi}}
\left[1 + \left(\frac{\sigma^2(M,z)}{a \delta_{cr}^2(z)}
\right)^p\right]\frac{\delta_{cr}(z)}{\sigma(M,z)}
\exp\left(-\frac{a \delta_{cr}^2(z)}{2 \sigma^2(M,z)}\right)\, .
\end{align}
We display redshift evolution of the SMT halo multiplicity function $\mathcal{F}_{\mathrm{SMT}}(M, z)$ as well as SMT-corrected DHMF $\frac{dn}{d \ln M}(M,z)\big\rfloor_{SMT}$ in Fig. \ref{f_SMT(z).pdf} and Fig. \ref{dnBYdlnM(z).pdf}, respectively for a typical mass $M=10^{13}\mathrm{h}^{-1}M_{\odot}$ with different model parameter $n$. We observe that both plots explicitly showcase a similar nature. In the low as well as high redshifts, they mimic the $\Lambda$CDM model. At lower redshifts, they grow rapidly, i.e, halo abundance forms in the late-time Universe, and their value increases with decreasing $n$. We also observe that apart from the background, they also depend on perturbative quantities like MDC and critical overdensity.
%%%%%%%%%%%%%%%%%%%%%%%%%%%%%%%%%%%%%%%%%%%%%%%%%%%%%%%%%%%
\begin{figure}[htbp!]
    \centering
    \includegraphics[width=10cm,height=8cm]{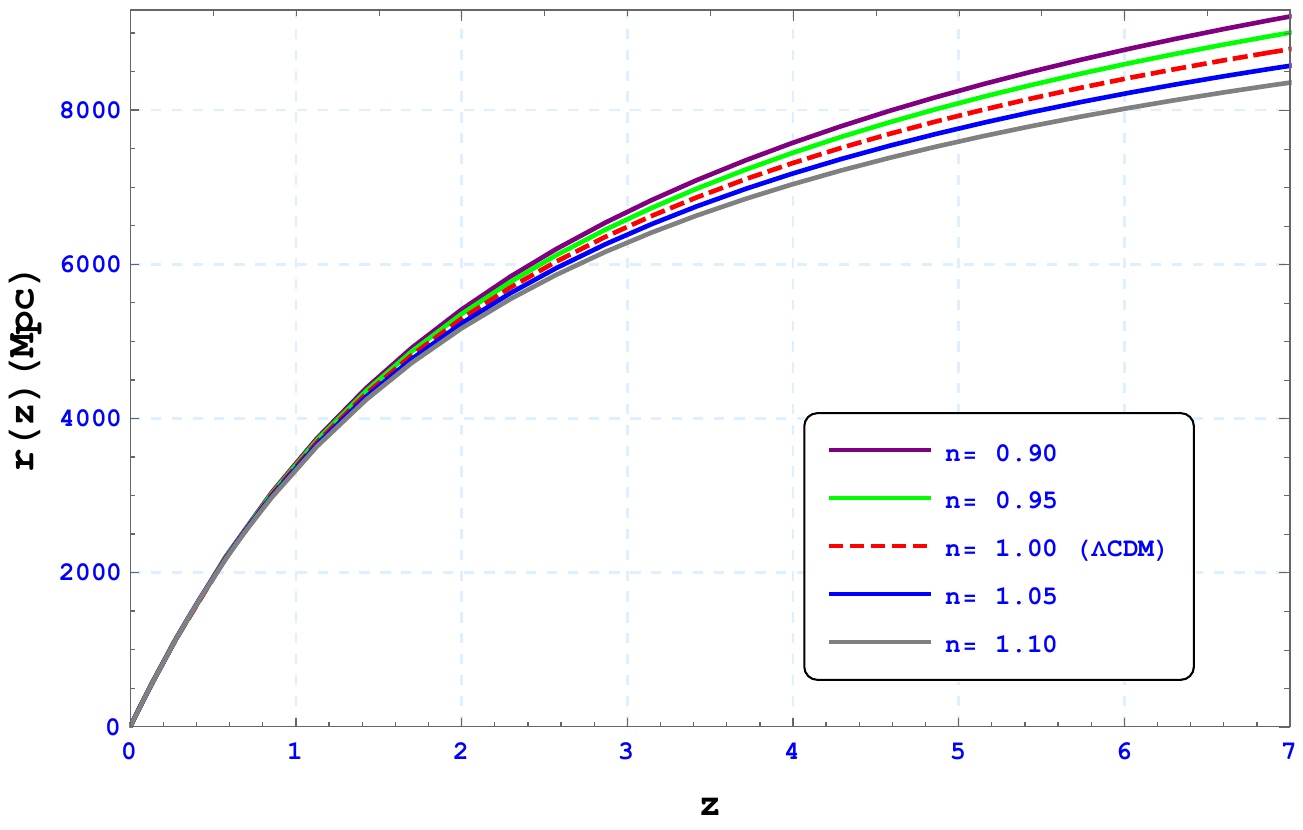}
    \caption{\small(Color online) Plot for comoving distance $r(z)$ versus redshift $z$  with different model parameter $n$}
    \label{r(z).pdf}
\end{figure}
%%%%%%%%%%%%%%%%%%%%%%%%%%%%%%%%%%%%%%%%%%%%%%%%%%%%%%%%%%%
\begin{figure}[htbp!]
    \centering
    \includegraphics[width=10cm,height=8cm]{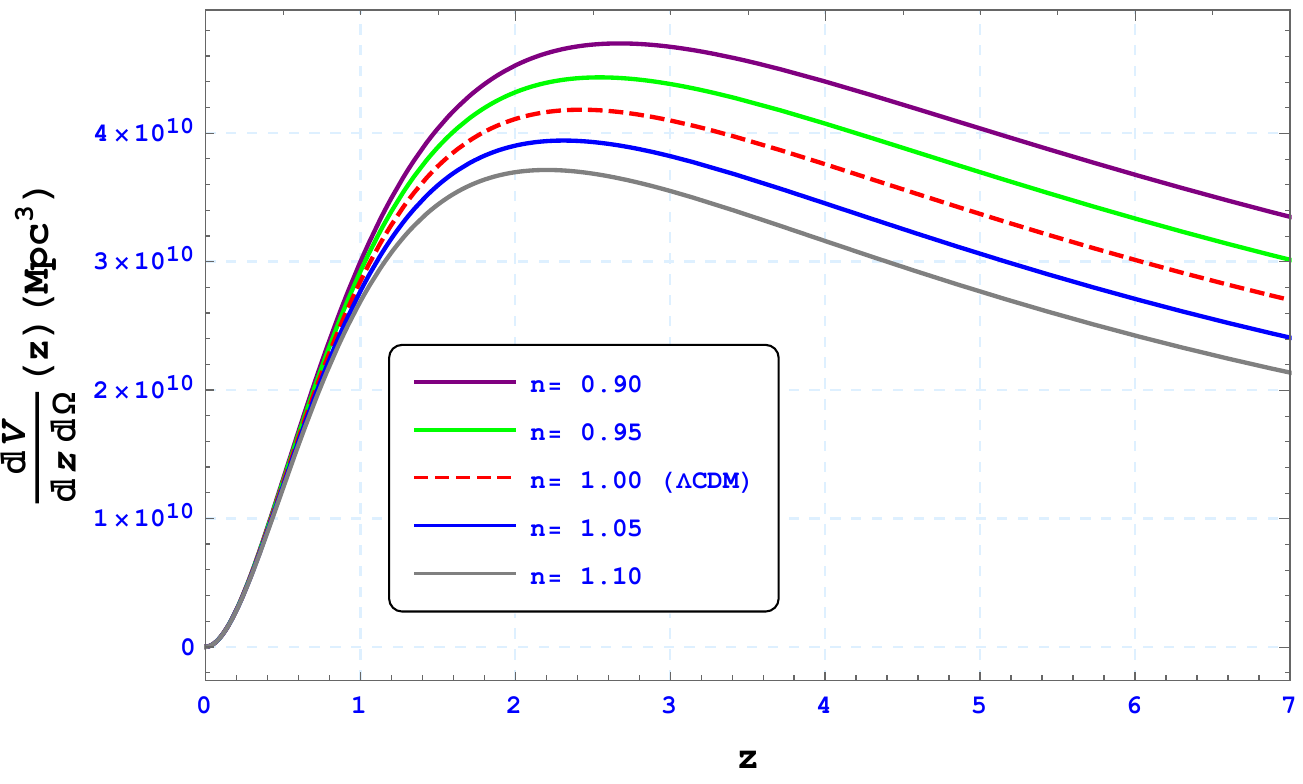}
    \caption{\small(Color online) Plot for comoving volume element $\frac{dV}{dz\,d\Omega}$ versus redshift $z$  with different model parameter $n$}
    \label{dVBYdzdOmega(z).pdf}
\end{figure}
%%%%%%%%%%%%%%%%%%%%%%%%%%%%%%%%%%%%%%%%%%%%%%%%%%%%%%%%%%%
%%%%%%%%%%%%%%%%%%%%%%%%%%%%%%%%%%%%%%%%%%%%%%%%%%%%%%%%%%%
\begin{figure*}[htbp!]
   \raisebox{0.8in}{($a$)}\includegraphics[width=10cm,height=8cm]{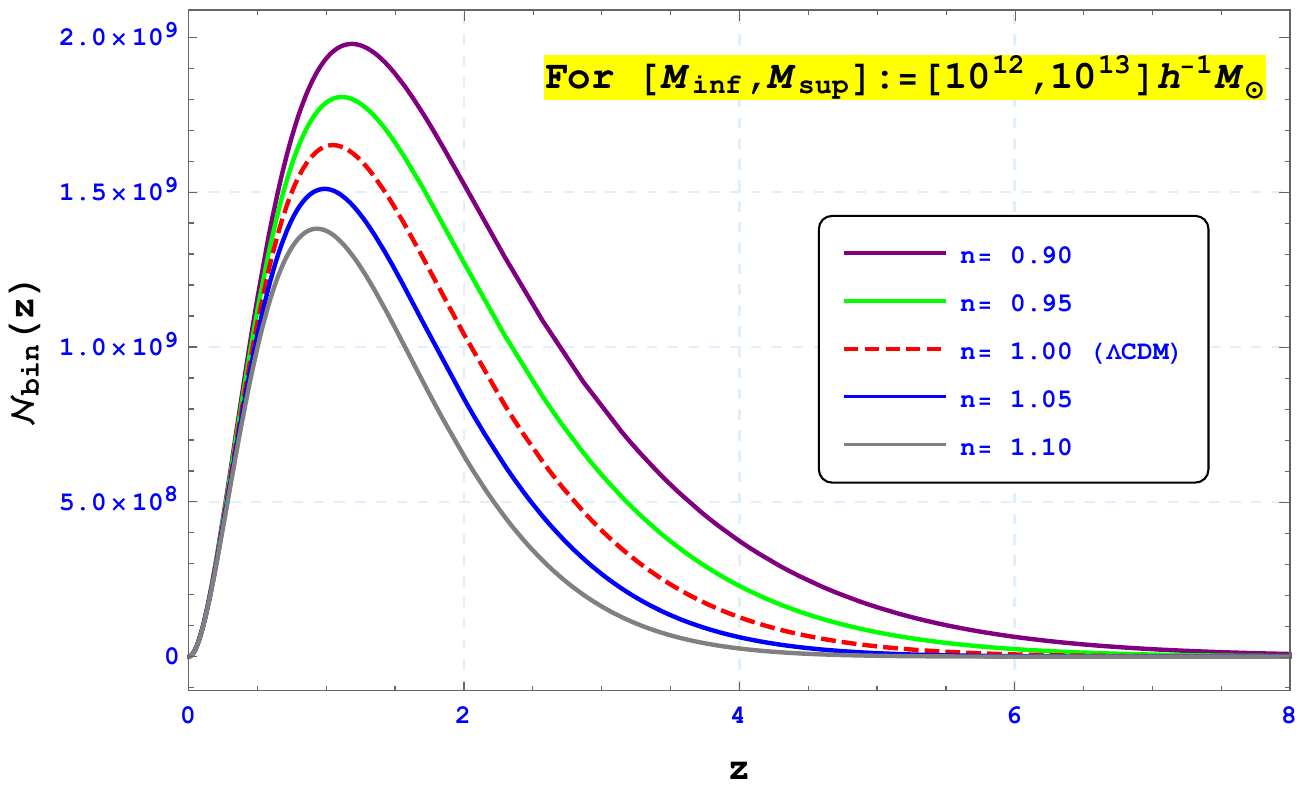}
   \hskip 0.03cm
  \raisebox{0.8in}{($b$)}\includegraphics[width=10cm,height=8cm]{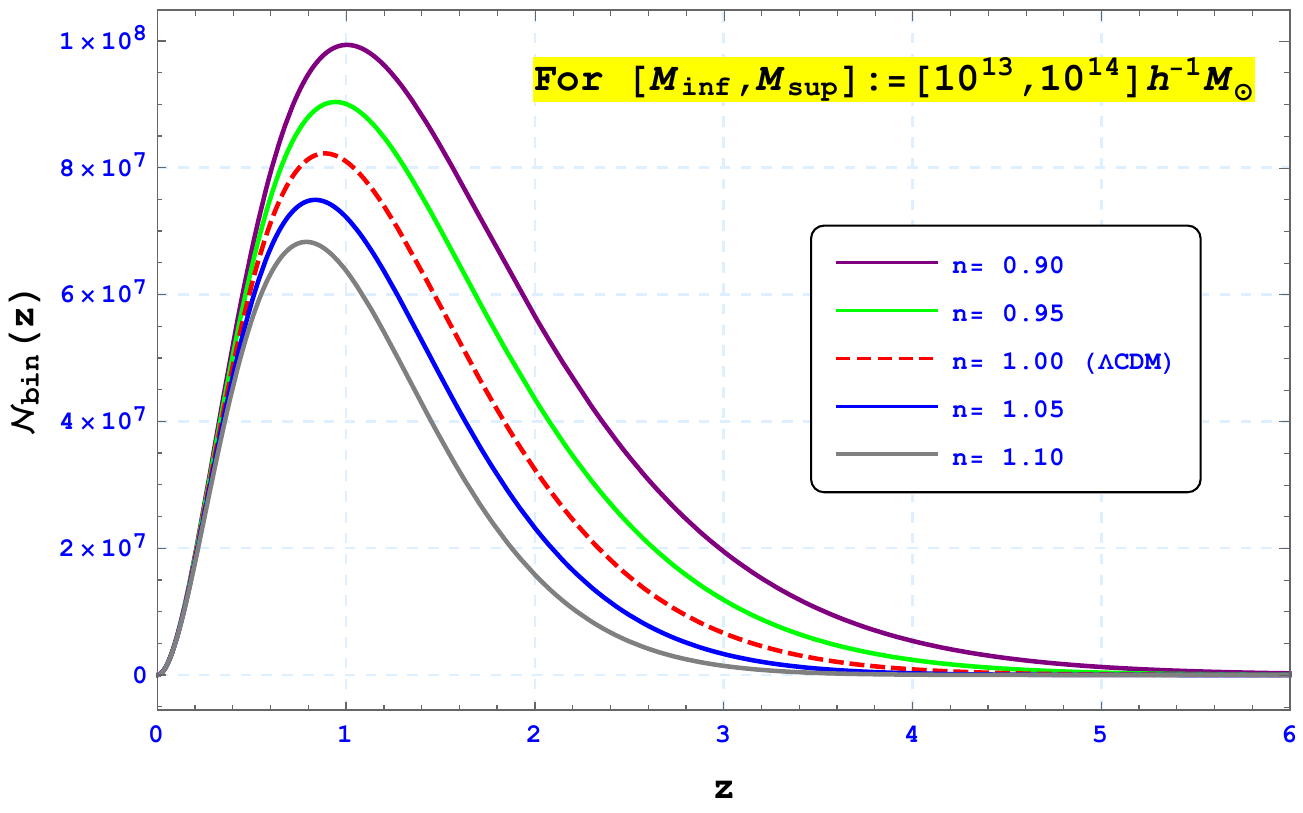}
  \caption{\small(Color online) Plots for halo number counts $\mathcal{N}_{bin}$ versus redshift $z$ for two specified mass bins with different model parameter $n$} 
 \label{N_bin plot}
\end{figure*}
%%%%%%%%%%%%%%%%%%%%%%%%%%%%%%%%%%%%%%%%%%%%%%%%%%%%%%%%%%%
\par Lastly, for the full sky survey of DM halo number count in the  GMHE-inspired modified cosmology, we apply the effective number of collapsed structures per unit of redshift, in a specified mass bin $M_{inf}<M<M_{sup}$ \cite{nunes2006structure}
\begin{equation}\label{N_bin}
    \mathcal{N}_{bin} \equiv \frac{dN}{dz} \overset{\text{def}}{=} \int_{4\pi}d\Omega\int_{M_{inf}}^{M_{sup}}\frac{1}{M}\frac{dn}{d \ln M}\frac{dV}{dz\,d\Omega}dM\,,
\end{equation}
where
\begin{equation}
    \frac{dV}{dz\,d\Omega} \overset{\text{def}}{=} \frac{\,c\, r^2(z)}{H(z)}
\end{equation}
is the elementary comoving volume and
\begin{equation}
    r(z)\overset{\text{def}}{=}c\int_{0}^{z}H^{-1}(\tilde{z})d\tilde{z}
\end{equation}  
denotes the comoving distance. The redshift evolution along with the dependence of model parameter $n$ of the comoving distance and the comoving volume element is displayed in Fig. \ref{r(z).pdf} and Fig. \ref{dVBYdzdOmega(z).pdf}, respectively. We see that the comoving distance shows a monotonic, concave-down growth curve as we look back in time. The elementary comoving volume first increases from $z=0$, attains a peak (around $1\lesssim z\lesssim3$, depending on model parameter $n$), and declines at higher redshifts. Both quantities increase with the decrement of the values of the model parameter $n$, and they are smoothly in agreement with the $\Lambda$CDM profile at the lower redshifts. It is also observed that they solely rely on the background expansion, rather than on the perturbative quantities. The halo number counts in mass bins in the GMHE-inspired modified cosmology, $\mathcal{N}_{bin}\equiv \frac{dN}{dz}$, can be obtained from Eq. (\ref{N_bin}), displayed in Fig. \ref{N_bin plot} for two different specified mass bins $[M_{inf}, M_{sup}]$ as $[10^{12}, 10^{13}]\mathrm{h}^{-1}M_\odot$ and $[10^{13}, 10^{14}]\mathrm{h}^{-1}M_\odot$. We note that $\mathcal{N}_{bin}$ decreases with increasing model parameter $n$, and a slight shift of the peak redshifts towards earlier time for structure formation is observed for various values of model parameter. At both ends of the redshift spectrum, all the curves converge asymptotically with the $\Lambda$CDM profile. As all the curves peak near the late-time (around $0.5\lesssim z \lesssim 1.5$, depending on model parameter), an excessive number of structures may form during these later epochs. We also see that heavier structures are deficient by comparing two mass bins. These align with the hierarchical model of large-scale structure formation, as confirmed by the works in Refs. \cite{abramo2007structure,liberato2006dark, farsi2022structure,farsi2023evolution}.
%%%%%%%%%%%%%%%%%%%%%%%%%%%%%%%%%%%%%%%%%%%%%%%%%%%%%%%%%%%%%%%%%%%%%%%%%%%%%%%%%%%%%%%%%%%%%%%%%%%%%%%%%%%%%%%%%%%
\section{Discussion and Conclusions} \label{Discussion and Conclusions}
\par From the recent generalization of SMHR through invoking the gravity-thermodynamics conjecture, it is found that the expansion history of the Universe can be altered in contrast to the fiducial $\Lambda$CDM scenario. In realistic phenomenological terms, one can visualize this as a potential extension to the standard $\Lambda$CDM cosmological framework. In particular, we have applied Friedmann dynamical equations, incorporating pressureless matter and a cosmological constant $\Lambda$ as DE, along with terms that include the GMHE-inspired modified cosmological model parameters. By considering this new entropic influence on the linear growth of overdensities, we have explored structure formation beyond the limitations of the $\Lambda$CDM paradigm employing a top-hat SC approach in conjunction with the SMT mass
function. 
\par Like the standard cosmological model, this modified cosmology establishes that the Universe underwent a deceleration-acceleration phase transition at around a redshift, $z_{\mathrm{tr}}\approx0.64$. Our investigation reveals that the inclusion of the $n$-term in the GMHE-inspired modified cosmological framework alters the transition epoch. It is shown that this model also satisfies the conditions under which the Universe attains thermodynamic equilibrium in the distant future. The nature of the various other cosmographic parameters, like jerk, snap, etc., also ascertains that the current Universe is accelerated. We discuss a new and well-established diagnostic approach to distinguish various cosmological models in relation to flat and non-flat $\Lambda$CDM frameworks, and find that the GMHE-inspired modified cosmology ($n\ne 1$) successfully passes all the litmus tests, thereby refuting both the flat and non-flat $\Lambda$CDM models. In this context, using the top-hat SC approach and modified Friedmann dynamical equations, an evolution equation of linear density perturbation is constructed and solved analytically. The linear MDC grows with decreasing redshift for all $n$. In this regard, we have investigated growth rates $f(z)$ and $f(z)\sigma_8(z)$ for the GMHE-inspired modified cosmology and find that suppression of growth near the lower redshifts, which is due to the dominance of the DE component. The shifting of $f(z)\sigma_8(z)$ peaks to lower redshifts for larger $n$ values indicates that the large-scale structures form at later stages in the GMHE-inspired modified cosmological scenario than in the standard $\Lambda$CDM counterpart for $n>1$; on the contrary, they form at an earlier epoch for $n<1$.  One prime endeavour of our present analysis was to inspect the DM halo/cluster survey in the case of GMHE-inspired modified cosmology. The modifications brought about by the GMHE-inspired modified cosmology influence the DM halo number counts, as the smaller values of $n$ give larger halo abundances and vice versa. It is also noticed that the halo mass function starts growing at lower redshifts, admitting halo structures form in later epochs. Besides, we observe that the halo number counts in mass bins, for which the less massive structures are more abundant, and they form at later epochs. These perfectly anticipate the hierarchical model of large-scale structure formation. 
\par Our present study shows that the  GMHE-inspired modified gravity scenario casts distinguishable imprints from the standard $\Lambda$CDM profile in all respects. It is tempting to ask whether the additional entropic effect could alleviate the $H_0$ and $\sigma_8$ tensions simultaneously and aid in its viability. A proper treatment for these tensions necessitates a comprehensive data analysis incorporating various available observational probes, such as Cosmic Microwave Background Radiation (CMBR), Supernovae Type Ia (SNe Ia), Baryon Acoustic Oscillations (BAO), Cosmic Chronometers (CC), $f\sigma_8$ growth rate, Hubble data, etc. On the observations related to halo abundance, there are experiments based on ongoing and upcoming observations, like the South Pole Telescope, eROSITA, etc. The South Pole Telescope survey, which relies on the Sunyaev–Zel'dovich effect detection, is expected to find a significant number of clusters and will be capable of accurately determining the number counts \cite{carlstrom2002cosmology}. The evolution of the cluster mass function reflects the linear growth of density fluctuations and can be used to constrain the model parameters of alternative gravity theories (See Ref. \cite{artis2024srg} for constraining model parameters in $f(\mathcal{R})$ gravity theory by examining the Hu-Sawicki parametrization alongside the first Spectrum Roentgen Gamma (SRG)/eROSITA All-Sky Survey (eRASS$1$) cluster catalog in the western Galactic hemisphere, together with the overlapping KiloDegree Survey, Dark Energy Survey Year-$3$, and Hyper Suprime-Cam data for weak lensing mass calibration.). Undoubtedly, these observations can serve as a powerful cosmological probe to substantiate the viability of the GMHE-inspired modified cosmological model and provide important insights into it. Furthermore, a study on nonlinear density perturbations for this modified model is also important and necessary in the context of structure formation, since the observational evidence \cite{huterer2015growth} indicates that the majority of the structures are formed from nonlinear evolution of overdensities in the dark age ($10 \lesssim z \lesssim 100$) \cite{miralda2003dark}. However, these topics are beyond the scope of our current study and will be explored in our future research.   
%%%%%%%%%%%%%%%%%%%%%%%%%%%%%%%%%%%%%%%%%%%%%%%%%%%%%%%%%%%%%%%%%%%%%%%%%%%%%%%%%%%%%%%%%%%%%%%%%%%%%%%%%%%%%%%%%
\acknowledgments
SM expresses gratitude to the Government of West Bengal, India, for providing the State-funded Senior Research Fellowship (SRF).
%%%%%%%%%%%%%%%%%%%%%%%%%%%%%%%%%%%%%%%%%%%%%%%%%%%%%%%%%%%%%%%%%%%%%%%%%%%%%%%%%%%%%%%%%%%%%%%%%%%%%%%%%%%%%%%
\section*{Data availability statement}
Data sharing does not apply to this article since no datasets were created or analyzed during this research.
%%%%%%%%%%%%%%%%%%%%%%%%%%%%%%%%%%%%%%%%%%%%%%%%%%%%%%%%%%%%%%%%%%%%%%%%%%%%%%%%%%%%%%%%%%%%
\section*{Code availability statement}
Code/Software sharing does not apply to this article, as the present study did not include the generation or analysis of any code or software.
%%%%%%%%%%%%%%%%%%%%%%%%%%%%%%%%%%%%%%%%%%%%%%%%%%%%%%%%%%%%%%%%%%%%%%%%%%%%%%%%%%%%%%%%%%%%
\section*{Conflict of interest}
The authors declare that they have no conflicts of interest to disclose.
%%%%%%%%%%%%%%%%%%%%%%%%%%%%%%%%%%%%%%%%%%%%%%%%%%%%%%%%%%%%%%%%%%%%%%%%%%%%%%%%%%%%%%%%%%%%%%%%%%%%%%%%%%%%%%%%%%%%%%%%%%%%%%%%%%%%
%\newpage
\bibliography{reference}

\end{document}